\documentclass[12pt]{article}
\usepackage{amsfonts, amsmath, amssymb, cite, graphicx, }

\usepackage{color}
\usepackage[all, knot]{xy}
\usepackage{tikz}
\usepackage{subfigure}
\usepackage{braket}

\usepackage[utf8]{inputenc}
\usepackage{ulem}

\usepackage{epstopdf}
\usepackage[footnotesize]{caption}
\usepackage{amsthm}
\usepackage{enumitem}
\usepackage{mathrsfs}
\usepackage{bm}

\usepackage[margin=3cm]{geometry}

\begin{document}

\begin{titlepage}
\vspace{0.5cm}
\begin{center}
{\Large \bf Quantum geometrical current and coherence of the open gravitation system: loop quantum gravity coupled with a thermal scalar field}

\lineskip .75em
\vskip 2.5cm
{\large Hong Wang$^{a}$, Jin Wang$^{b,}$\footnote{jin.wang.1@stonybrook.edu} }
\vskip 2.5em
 {\normalsize\it $^{a}$State Key Laboratory of Electroanalytical Chemistry, Changchun Institute of Applied Chemistry, Chinese Academy of Sciences, Changchun 130022, China\\
 $^{b}$Department of Chemistry and Department of Physics and Astronomy, State University of New York at Stony Brook, NY 11794, USA}
\vskip 3.0em
\end{center}
\begin{abstract}
Open quantum systems interacting with the environments often show interesting behaviors, such as decoherence, non-unitary evolution, dissipation, etc. It is interesting but still challenging to study the open quantum gravitation system interacting with the environments. In this work, we develop a general parameterized theoretical framework for the open quantum gravitation system.  We studied a specific model where a real scalar field plays the role of the environment and the spacetime is assumed to be homogeneous and isotropic. We quantize the spacetime through the loop quantum gravity. We show that if the scalar field is in the thermal equilibrium state, the spacetime geometry will reach the equilibrium state after the transient relaxation. For the non-steady state, the quantum geometry current emerges. We point out that the quantum geometry current and the coherence can together drive the evolution of the spacetime geometry. This provides us a new view on the evolution of the spacetime geometry. Our results show that the coherence of the spacetime monotonically decreases as the temperature of the bath decreases. It helps the understanding of how a classical cold universe can emerge from an initial hot quantum universe.
  \end{abstract}
\end{titlepage}

\baselineskip=0.7cm

\tableofcontents
\newpage
\section{Introduction}

In quantum mechanics, the evolution of any isolated system can be described by the Schr\"{o}dinger equation.
The evolution described by the schr\"{o}dinger equation is unitary, thus time reversible.
However, the open system coupled with environments is  more common in reality. The open system usually can not be described by the schr\"{o}dinger equation. But it can be described by the mixed density matrix $\rho=\sum_{i}\omega_{i}|\Psi_{i}\rangle \langle\Psi_{i}|$. The evolution of the open system can be described by the von Neumann equation~\cite{HF}
\begin{equation}
\label{eq:1.1}
 \frac{d\rho}{dt}=-i\mathrm{Tr}_{B}[\hat{H},\rho_{tot}].
\end{equation}
The trace is taken over the environment. $\rho_{tot}$ represents the density matrix of the total system. $\rho$ is the reduced density matrix of the open system, and $\hat{H}$ is the Hamiltonian operator of the total system (system + environment). These quantities in equation \eqref{eq:1.1} are all defined in the schr\"{o}dinger picture. Throughout this study, we use natural units $\hbar=c=G=k_{B}=1$ and the signature (-,+,+,+).

In general, equation \eqref{eq:1.1} is very difficult to solve. However, if the environment is a large heat bath with many degrees of freedom, and the coupling between the system and the environment is very week, one can approximately treat the density matrix $R_{0}$ of the environment as being not changed with time~\cite{HF,HC}. By using the Born-Markov approximation, the equation \eqref{eq:1.1} becomes~\cite{HF,HC}
\begin{equation}
\label{eq:1.2}
\dot{\tilde{\rho}}=-\int_{0}^{\infty}dt'\mathrm{Tr}_{B}\Big\{[\tilde{H}_{int}(t),[\tilde{H}_{int}(t-t'),\tilde{\rho}(t)R_{0}]]\Big\}.
\end{equation}
Equation \eqref{eq:1.2} is the so called quantum master equation. Here, $\tilde{\rho}$ is the density matrix of the system, and $\tilde{H}_{int}$ is the interaction Hamiltonian operator. All these operators are defined in the interaction picture. The evolution of the system described by equation \eqref{eq:1.2} can be non-unitary. The entanglement entropy and other quantum correlations as well as the
energy of the system can vary with time. Thus the evolution may be irreversible. Decoherence may occur due to the coupling between the system and the environment~\cite{MS}.

In the very early time of the universe, one expects that the quantum effect of the gravitation system should be important. Such as the quantum fluctuation of the metric. For the present time, the quantum effect of the metric has not been observed in almost all the experiments. This may be due to the decoherence occurred in the gravitation system~\cite{JJ,TP,AAC,NA,AB}. On the other hand, black hole has the Hawking radiation, so the entropy of the black hole can change with time due to the decrease of the event horizon in time. The evolution of the black hole (not including the radiation field and another environment) maybe non-unitary. Both these features are difficult to be described by the Schr\"{o}dinger type equations. Instead, they should be studied in the framework of the open quantum gravitation system.

The quantum master equation approach for the density matrix evolution for the usual open quantum system can be used to study the open quantum gravitation system. Assuming that our universe can be seen as an isolated system, and we can use a pure state to describe our universe. The general covariance principle requires the Hamiltonian $H_{c}$ of our universe to be zero~\cite{MC,CK,CR}. If we canonically quantize the Hamiltonian $H_{c}$, then we can obtain the Wheeler-DeWitt equation~\cite{CK,CR}
\begin{equation}
\label{eq:1.3}
\hat{H}_{c}|\varphi\rangle_{u}=0.
\end{equation}
Here $|\varphi\rangle_{u}$ is the wave function of the universe. However, it is difficult to understand the evolution of our universe based on the Wheeler-DeWitt equation \eqref{eq:1.3}~\cite{TB,MY,SV,AK}, due to the time derivative of the wave function of the universe being zero.

We consider the Hamiltonian of our universe to be composed of certain parts $H_{c}=H_{grav}+H_{B} +H_{int}$, where $H_{grav}$  and $H_{B}$ represent the Hamiltonian of the gravitation system and the environment, respectively. And $H_{int}$ is the interaction Hamiltonian between the gravitation system and the environment.  Although the covariance principle requires the total Hamiltonian to be zero~\cite{CK}, $H_{c}=0$, it does not require the Hamiltonian of the subsystem to be equal to zero. Therefore, if there are other matter fields, the Hamiltonian of the gravitation system $H_{grav}\neq 0$. If we focus on the evolution of the gravitation system $H_{grav}$, it should be seen as an open system. We should describe it by the mixed density matrix $\rho$. Although the covariance principle requires $d|\varphi\rangle_{u}/dt=0$, for the gravitation system, in general $d\rho/dt\neq 0$.

For the usual canonical quantum cosmology, the case for the gravity minimally coupled with the scalar field has been extensively studied. For the case of non-minimal coupling, there are only a few studies on the  influence of the scalar field to the tunneling of the universe~\cite{SWM,RF}. There have been several studies which treated the quantum spacetime as an open system and studied the associated features, such as the decoherence, the evolution of the spacetime geometry, etc~\cite{AB,NA}. In loop quantum cosmology, there were also some studies on the influence of the scalar field to the spacetime structure, see~\cite{AAP,AE} and the references therein. In these studies, the scalar field was assumed to only depend on the coordinate time and can be seen as a time parameter. In this case, the degree of freedom of the scalar field is one. The role of the scalar field is equivalent to the time variable~\cite{AAP,AE}. However, even if the spacetime is homogenous, the scalar field can also depend on both the time coordinate and the space coordinate via the phase factor~\cite{LP}. When the scalar field depends on both the space and the time coordinates, it usually has infinite vibrational modes, that is, the degrees of freedom are infinite. We take an initial step to treat the loop quantum spacetime as an open quantum system and study the influence of thermal scalar field on the spacetime structure while the scalar field depend both on the time coordinate and the space coordinate.

In this work, we develop a general parameterized theoretical framework for the open quantum gravitation system. This framework also holds for the non-inertial frame.
We studied a specific model where the real scalar field plays the role of the environment and the spacetime is homogeneous and isotropic. The goal of this model is not to describe the exact quantum evolution of our physical universe (This is important but often very difficult), but still reveal some interesting features.

We quantize the gravitation system through the loop quantum gravity.  In loop quantum gravity, all the eigenvalues of the volume operator form a discrete spectrum. This can be used to simplify our calculations. We assume that the scalar field to be in the thermal equilibrium, then the scalar field can induce the non-unitary evolution of the gravitation system. After a period of times, the gravitation system will reach the quasi-steady state. In this quasi-steady state, the spacetime geometry is in the equilibrium state where the detailed balance is preserved. For the non-steady state, the non-zero quantum geometry current emerges. This current and the coherence can drive the evolution of the spacetime. This is consistent with the consensus in the quantum thermodynamics that we can extract work from the coherence~\cite{XZG,KMJ}. We found that the coherence monotonically decreases as the temperature of the bath decreases. This indicates that a cold classical universe can emerge from an initial hot quantum universe. After taking the continuous limit, we obtain a transition probability of the spacetime from certain initial state to the eigenstate of the volume operator.

\section{Parameterized theory for the open quantum gravitation system}
\subsection{Parameterized real scalar field theory}
To clearly present our approach, let us review the main results of the parameterized real scalar field theory. The action of the real scalar field $\phi$ in the 4-dimensional Minkowski spacetime is given as~\cite{CK}
\begin{equation}
\label{eq:2.1}
\mathcal{S}=\int dX^{4} \mathcal{L}(\phi,\frac{\partial\phi}{\partial X^{\mu}}).
\end{equation}
Here,  $\mathcal{L}$ is the Lagrangian density of the scalar field, and  $X^{\mu}$ is the Cartesian coordinates of the 4-dimensional Minkowski spacetime. $X^{0}$ is the time coordinate, and $X^{a}$ ($a=1,2,3$) is the space coordinates. In these coordinates, the metric of the Minkowski spacetime is $ds^{2}=\eta_{\mu\nu}dX^{\mu}dX^{\nu}$ with $\eta_{\mu\nu}=diag(-,+,+,+)$. In this work, we set the Latin indexes a, b,c...=1, 2, 3 and the Greek indexes $\mu$, $\nu$...=0, 1, 2, 3.

We can generalize the above to arbitrary coordinates. Thus in arbitrary reference frame:
\begin{equation}
\label{eq:2.2}
X^{\mu}=X^{\mu}(x^{\nu}), \qquad  \mu,\nu=0,1,2,3.
\end{equation}
 In the coordinate system $x^{\mu}$, the spacetime metric is $ds^{2}=g_{\mu\nu}dx^{\mu}dx^{\nu}$, set $X^{\alpha}_{,\mu}=\partial X^{\alpha}/\partial x^{\mu}$, then $g_{\mu\nu}=\eta_{\alpha\beta}X^{\alpha}_{,\mu} X^{\beta}_{,\nu}$. We will always constrain that $g_{00}<0$ and $g_{aa}>0$ ($a=1,2,3$). Under this coordinate transformation, the scalar field is invariant, so the action \eqref{eq:2.1} becomes
\begin{equation}
\label{eq:2.3}
\mathcal{S}=\int dx^{4} J \mathcal{L}(\phi,\frac{\partial\phi}{\partial x^{\nu}}\cdot \frac{\partial x^{\nu}}{\partial X^{\mu}})\equiv \int dx^{4} \mathcal{\tilde{L}},
\end{equation}
where $\mathcal{\tilde{L}}=J \mathcal{L}$ is the Lagrangian density of the scalar field in the reference frame $x^{\nu}$, and $J$ is the Jacobian  determinant,
\begin{equation}
\label{eq:2.4}
J\equiv \frac{\partial (X^{0},X^{1},X^{2},X^{3})}{\partial (x^{0},x^{1},x^{2},x^{3})}=\epsilon_{\mu\nu\sigma\delta}\frac{\partial X^{\mu}}{\partial x^{0}}\frac{\partial X^{\nu}}{\partial x^{1}}\frac{\partial X^{\sigma}}{\partial x^{2}}\frac{\partial X^{\delta}}{\partial x^{3}}.
\end{equation}
Here, $\epsilon_{\mu\nu\sigma\delta}$ is the signature of the permutation of (0123). If any two indices are equal to each other, then $\epsilon_{\mu\nu\sigma\delta}=0$. For example, $\epsilon_{0123}=1$, $\epsilon_{1023}=-1$, $\epsilon_{0023}=0$... If the coordinate $x^{\nu}$ is related to a non-inertial observer, then for this observer, the dynamics of the scalar field $\phi$ should be described by the Lagrangian $\mathcal{\tilde{L}}$.

The conjugate momentum for the scalar field $\phi$ in the coordinate $x^{\mu}$ is $\tilde{P}_{\phi}=\partial \mathcal{\tilde{L}}/\partial \dot{\phi}$, here, $\dot{\phi}=\partial \phi/\partial x^{0}$. Then we can obtain the Hamiltonian of the scalar field $\phi$ in the coordinates $x^{\mu}$ as
\begin{equation}
\label{eq:2.6}
\mathcal{\tilde{H}}=\tilde{P}_{\phi}\dot{\phi}-\mathcal{\tilde{L}}
                   =J\frac{\partial x^{0}}{\partial X^{\mu}}\Big(\frac{\partial \mathcal{L}}{\partial (\partial \phi/\partial X^{\mu})}\frac{\partial \phi}{\partial X^{\nu}}-\delta^{\mu}_{\nu}\mathcal{L}\Big)\dot{X}^{\nu},
\end{equation}
where $\dot{X}^{\nu}=\partial X^{\nu}/\partial x^{0}$. Noted that the canonical energy-momentum tensor of the scalar field is given by
\begin{equation}
\label{eq:2.7}
T^{\mu}_{\nu}=\frac{\partial \mathcal{L}}{\partial (\partial \phi/\partial X^{\mu})}\frac{\partial \phi}{\partial X^{\nu}}-\delta^{\mu}_{\nu}\mathcal{L}.
\end{equation}
For the scalar field in the Minkowski spacetime, the canonical energy-momentum tensor is symmetric and is equivalent to the Belinfante-Rosenfeld tensor~\cite{MR}. For other situations (For curved spacetime or when spin of the field is not zero), the canonical energy-momentum tensor is usually not symmetric. So the Hamiltonian \eqref{eq:2.6} can be written as
\begin{equation}
\label{eq:2.8}
\mathcal{\tilde{H}}=JT^{\mu}_{\nu}\dot{X}^{\nu}\frac{\partial x^{0}}{\partial X^{\mu}}.
\end{equation}

On the other hand, in the Lagrangian $\mathcal{\tilde{L}}$, we can also treat $X^{\mu}(x)$  as a canonical variable~\cite{CK}. Its conjugate momentum is given by
\begin{equation}
\label{eq:2.9}
\Pi_{\mu}=\frac{\partial \mathcal{\tilde{L}}}{\partial (\partial X^{\mu}/\partial x^{0})}
        =J\frac{\partial x^{0}}{\partial X^{\mu}}\mathcal{L}-J\frac{\partial x^{0}}{\partial X^{\nu}}\frac{\partial \mathcal{L}}{\partial (\partial \phi/\partial X^{\nu})}\frac{\partial \phi}{\partial X^{\mu}}.
\end{equation}
Combining equation \eqref{eq:2.7} and \eqref{eq:2.9}, one can obtain
\begin{equation}
\label{eq:2.10}
\Pi_{\mu}=-J\frac{\partial x^{0}}{\partial X^{\nu}}T^{\nu}_{\mu}.
\end{equation}
Bringing equation \eqref{eq:2.10} into \eqref{eq:2.8}, we have
\begin{equation}
\label{eq:2.10a}
\mathcal{\tilde{H}}=JT^{\mu}_{\nu}\dot{X}^{\nu}\frac{\partial x^{0}}{\partial X^{\mu}}=-\Pi_{\nu}\dot{X}^{\nu}.
\end{equation}
Equation \eqref{eq:2.10a} indicates that
\begin{equation}
\label{eq:2.10b}
\big(\Pi_{\nu}+JT^{\mu}_{\nu}\frac{\partial x^{0}}{\partial X^{\mu}}\big)\dot{X}^{\nu}=0.
\end{equation}

If we define
\begin{equation}
\label{eq:2.11}
\mathcal{H}_{\mu}=\Pi_{\mu}+J\frac{\partial x^{0}}{\partial X^{\nu}}T^{\nu}_{\mu},
\end{equation}
then, equation \eqref{eq:2.10b} can be rewritten as $\mathcal{H}_{\mu}\dot{X}^{\mu}=0$. This should be valid for arbitrary coordinates transformation \eqref{eq:2.2}, this means
\begin{equation}
\label{eq:2.12}
\mathcal{H}_{\mu}=0.
\end{equation}
In addition, combining equation \eqref{eq:2.10a} and \eqref{eq:2.11}, one can show that the relation between $\mathcal{\tilde{H}}$ and $\mathcal{H}_{\mu}$ is
\begin{equation}
\label{eq:2.12a}
\mathcal{\tilde{H}}=\dot{X}^{\mu}(\mathcal{H}_{\mu}-\Pi_{\mu}).
\end{equation}
For convenience, one can call the quantity $\mathcal{H}_{\mu}$ the Super-Hamiltonian vector. We should point out that in \eqref{eq:2.12}, $\mathcal{H}_{\mu}$ is weakly equal to zero, i.e. equal to zero on the physical configurations. In fact, $\mathcal{H}_{\mu}=0$ is the result of the general covariance of the scalar field theory.

Usually, one can decompose \eqref {eq:2.12} into the components~\cite{CK} (3+1 decomposition):
\begin{eqnarray}\begin{split}
\label{eq:2.13}
\mathcal{H}_{\bot}=\mathcal{H}_{\nu}n^{\nu}=0,\\
\mathcal{H}_{a}=\mathcal{H}_{\nu}X^{\nu}_{,a}=0.
\end{split}
\end{eqnarray}
Here, $n^{\nu}$ is the normal vector orthogonal to the hypersurface $x^{0}=constant$, and $X^{\nu}_{,a}=\partial X^{\nu}/\partial x^{a}$ are the tangential vectors parallel to the hypersurface $x^{0}=constant$. $\mathcal{H}_{\bot}=0$ is called the Hamiltonian constraint and $\mathcal{H}_{a}=0$ are called the diffeomorphism constraints. $\mathcal{H}_{a}$ are the infinitesimal generators of the diffeomorphism transformation.

Carrying out the canonical quantization procedure to the scalar field, then the Schr\"{o}dinger equation becomes
\begin{equation}
\label{eq:2.18}
i\frac{\delta |\Psi(\phi(x),X^{\mu}(x))\rangle}{\delta X^{\mu}(x)}=\mathcal{\hat{H}}_{\mu}|\Psi(\phi(x),X^{\mu}(x))\rangle=0
\end{equation}
with the constraint \eqref{eq:2.12}. For the ordinary Schr\"{o}dinger equation $i\partial \psi/\partial t=\hat{H}\psi$, the wave function is evolved along the time parameter ``$t$". Similarly, for equation \eqref{eq:2.18}, if $\mathcal{\hat{H}}_{\mu}|\Psi(\phi(x),X^{\mu}(x))\rangle\neq 0$, then its solution can be evolved along any family of hypersurfaces which is described by $X^{\mu}(x)$. Because of this, the coordinates $X^{\mu}(x)$ are also called the bubble-time~\cite{CK}. We can define the density matrix as $\rho=|\Psi(\phi(x),X^{\mu}(x))\rangle\langle\Psi(\phi(x),X^{\mu}(x))|$, the equation \eqref{eq:2.18} can be written as another form:
\begin{equation}
\label{eq:2.19}
\frac{\delta \rho}{\delta X^{\mu}(x)}=-i[\mathcal{\hat{H}}_{\mu},\rho]=0.
\end{equation}

Equations \eqref{eq:2.18} and \eqref{eq:2.19} are equivalent to each other when the scalar field is an isolated system. But when the scalar field is an open system, there are other matter fields interacting with it. The scalar field maybe entangled with the environment, and the entanglement entropy of the scalar field may change in time. Thus, the evolution of the scalar field can be non-unitary. In this case, equations \eqref{eq:2.18} can not be straightforwardly used to describe the evolution of the scalar field. For the total system composed by the scalar field and the environment,
\begin{equation}
\label{eq:2.20}
\mathcal{\hat{H}}_{\mu,tot}=\mathcal{\hat{H}}_{\mu,sca}+\mathcal{\hat{H}}_{\mu,env}+\mathcal{\hat{H}}_{\mu,int}.
\end{equation}
Here, $\mathcal{\hat{H}}_{\mu,tot}$ is the Super-Hamiltonian vector operator of the total system, $\mathcal{\hat{H}}_{\mu,sca}$ is the Super-Hamiltonian vector operator of the scalar field, $\mathcal{\hat{H}}_{\mu,env}$ is the Super-Hamiltonian vector operator of the environment, and $\mathcal{\hat{H}}_{\mu,int}$ is the interaction Super-Hamiltonian vector operator between the scalar field and the environment. For the total system (an isolated system), we can describe it by the equations $-i[\mathcal{\hat{H}}_{\mu,tot},\rho_{tot}]=0$. But for the scalar field,  it is an open system, we should describe its evolution by the following equations
\begin{equation}
\label{eq:2.21}
\frac{\delta \rho}{\delta X^{\mu}(x)}=-i\mathrm{Tr}_{B}[\mathcal{\hat{H}}_{\mu,tot}\,,\rho_{tot}],
\end{equation}
where the trace is taken over the environment. Equations \eqref{eq:2.21} are the parameterized version of the von Neumann equation. Although for the total system, the covariance requires that $[\mathcal{\hat{H}}_{\mu,tot}\,, \rho_{tot}]=0$. Thus $\delta \rho_{tot}/\delta X^{\mu}(x)=0$. For the scalar field as a subsystem, the general covariance does not require its Super-Hamiltonian vector to be equal to zero. In general, $\delta \rho/\delta X^{\mu}(x)\neq 0$ for the subsystem.

Equations \eqref{eq:2.18} or \eqref{eq:2.19} are the foundations of the parameterized theory for the isolated real scalar field.  If the scalar field is an open system, the foundations of the parameterized theory for the scalar field should be given by the equations \eqref{eq:2.21}. Equations \eqref{eq:2.21} can describe certain interesting features of the scalar field as an open system, such as the decoherence, the non-unitary evolutions, and the nonequilibrium evolutions. We pointed out that all the results in this subsection are derived in the Minkowski spacetime. We will generalize to the curved spacetime in the next.

\subsection{Parameterized theory for the gravitation system }
In this work, we just focus on the open quantum gravitation system. For the gravitational field coupled with a scalar field, The total action is given as
\begin{equation}
\label{eq:2.22}
\mathcal{S}=\frac{1}{16\pi}\int dx^{4}\sqrt{-g}R +\mathcal{S}_{\phi}(\phi,g_{\mu\nu}).
\end{equation}
Here, $g$ is the determinant of the 4-dimensional spacetime metric $g_{\mu\nu}$, and $R$ is the Ricci scalar. On the right hand side of equation \eqref{eq:2.22}, the first term is the Einstein-Hilbert action of the spacetime. And the second term represents the action of the scalar field in the curved spacetime.

In the curved spacetime, the symmetric energy-momentum tensor of the scalar field is defined as~\cite{LP}
\begin{equation}
\label{eq:2.13a}
\Theta_{\mu\nu}\equiv-\frac{2}{\sqrt{-g}}\frac{\delta \mathcal{S}_{\phi}(\phi,g_{\mu\nu})}{\delta g^{\mu\nu}}.
\end{equation}
In \eqref{eq:2.13a}, we use $\mathcal{S}_{\phi}(\phi,g_{\mu\nu})$ to represents the action of the scalar field in the curved spacetime. In curved spacetime, for the scalar field,  $\nabla^{\mu}\Theta_{\mu\nu}=0$, and $\Theta_{\mu\nu}$ is different from the canonical energy momentum tensor $T_{\mu\nu}$ ($T_{\mu\nu}=-g_{\mu\nu}\mathcal{L}+\nabla_{\nu}\phi(\partial \mathcal{L}/\partial\nabla^{\mu}\phi)$)~\cite{LP}.

In order to quantize the system by the canonical quantization scheme, we need to carry out the so-called 3+1 decomposition for it, i.e., foliating the 4-dimensional spacetime into a family of 3-dimensional space-like hypersurfaces $\sum_{t}$~\cite{CK,FO}. Different hypersurfaces are distinguished by the time coordinate $t$ ($t=x^{0}$). In each hypersurface, the time coordinate $t$ is the same on different places, $\sum_{t}$ is a family of Cauchy surfaces. We express this family of hypersurface $\sum_{t}$ by
\begin{equation}
\label{eq:2.23}
Y^{\mu}=Y^{\mu}(t,x^{a}).
\end{equation}
Here, $Y^{\mu}$ are the coordinates of the embedded space.

Defining $h_{ab}$ as the metric of the hypersurface $\sum_{t}$,  set $Y^{\mu}_{,a}=\partial Y^{\mu}/\partial x^{a}$, the relation between $h_{ab}$ and $g_{\mu\nu}$ is $h_{ab}=g_{\mu\nu}Y^{\mu}_{,a}Y^{\nu}_{,b}$. We can treat $h_{ab}$ as a canonical variable, the corresponding conjugate momentum is given by~\cite{CK}
\begin{equation}
\label{eq:2.24}
P^{ab}=\frac{\partial \mathcal{L}}{\partial \dot{h}_{ab}}=\frac{1}{16\pi}G^{abcd}K_{cd}.
\end{equation}
Here, $h$ is the determinant of the metric $h_{ab}$ and $K_{ab}$ is the extrinsic curvature of the hypersurface $\sum_{t}$. $G^{abcd}\equiv\sqrt{h}(h^{ac}h^{bd}+h^{ad}h^{bc}-2h^{ab}h^{cd})/2$ represents the DeWitt metric. Note that the Latin alphabet indices $a,b,c,d$ are contracted with the metric $h_{ab}$, and the Greek alphabet indices $\mu,\nu$ are contracted with the metric $g_{\mu\nu}$.

After carrying out the 3+1 decomposition, analogous to the equations \eqref{eq:2.13}, we obtain the following constraint equations for the system~\cite{CK}:
\begin{equation}
\label{eq:2.28}
\mathcal{H}_{\bot}=\mathcal{H}_{\nu}\,n^{\nu}=16\pi G_{abcd}P^{ab}P^{cd}-\frac{\sqrt{h}}{16\pi}R^{(3)}+\sqrt{h}\Theta_{\mu\nu}n^{\mu}n^{\nu}=0,
\end{equation}
\begin{equation}
\label{eq:2.29}
\mathcal{H}_{a}=\mathcal{H}_{\nu}Y^{\nu}_{,a}=-2\nabla_{b}P_{a}^{b}+\sqrt{h}h^{\mu}_{a}\Theta_{\mu\nu}n^{\nu}=0.
\end{equation}
Here, $R^{(3)}$ is the Ricci scalar of the hypersurface $\sum_{t}$, $Y^{\mu}_{,a}$ are the tangential vectors parallel to the hypersurface $\sum_{t}$, and $\nabla_{b}$ is the covariant derivative with respect to the metric $h_{ab}$. Equation \eqref{eq:2.28} is the Hamiltonian constraint, and equations \eqref{eq:2.29} are the diffeomorphism  constraints (also called momentum constraint). These equations are the foundations of the parameterized theory  for the gravitational field coupled with the scalar field. They completely determined the classical features of the system.

The quantum version of the classical constraint equations \eqref{eq:2.28} and \eqref{eq:2.29} are
\begin{equation}
\label{eq:2.31}
\mathcal{\hat{H}}_{\bot}|\Psi\rangle=0
\end{equation}
and
\begin{equation}
\label{eq:2.32}
\mathcal{\hat{H}}_{a}|\Psi\rangle=0,
\end{equation}
respectively. Equation \eqref{eq:2.31} is the Wheeler-DeWitt equation. Due to $\mathcal{\hat{H}}_{\mu}=(\mathcal{\hat{H}}_{\bot}\,,\mathcal{\hat{H}}_{a})$, we can write equations \eqref{eq:2.31} and \eqref{eq:2.32} in a more compact way: $\mathcal{\hat{H}}_{\mu}|\Psi\rangle=0$, or equivalently in the form of the parameterized von Neumann equation
\begin{equation}
\label{eq:2.33}
\frac{\delta \rho(g_{\mu\nu},\phi)}{\delta Y^{\mu}(x)}=-i[\mathcal{\hat{H}}_{\mu}\,,\rho(g_{\mu\nu},\phi)]=0
\end{equation}
with the constraints \eqref{eq:2.31} and \eqref{eq:2.32}. In equation \eqref{eq:2.33}, $\rho(g_{\mu\nu},\phi)$ represents the density matrix of the system.

If we just care about the information of the quantum spacetime, then we can treat the gravitational field and the scalar field as the open subsystem and the environment, respectively. After tracing out the degrees of freedom of the environment, one can obtain
\begin{equation}
\label{eq:2.34}
\frac{\delta \rho(g_{\mu\nu})}{\delta Y^{\mu}(x)}=-i\mathrm{Tr}_{B}[\mathcal{\hat{H}}_{\mu}\,,\rho(g_{\mu\nu},\phi)]
\end{equation}
where $ \rho(g_{\mu\nu})$ represents the reduced density matrix of the gravitational field. Noted that in this case, the gravitational field is not an isolated system, thus the general covariance principle does not require that the Super-Hamiltonian vector of the gravitational field is equal to zero. Therefore, $\delta \rho(g_{\mu\nu})/\delta Y^{\mu}(x)\neq0$. This is similar to the case of equation \eqref{eq:2.21}.

There are the similar results in more general situations. Generally, if the environment is composed of various matter fields. Assuming  $\mathcal{\hat{H}}_{\mu,tot}$ represents the Super-Hamiltonian vector operators of the total system. In general, $\mathcal{\hat{H}}_{\mu,tot}$ can be written as $\mathcal{\hat{H}}_{\mu,tot}=\mathcal{\hat{H}}_{\mu,grav}+\mathcal{\hat{H}}_{\mu,oth}$, where $\mathcal{\hat{H}}_{\mu,oth}$ represents the Super-Hamiltonian vector operator of the rest part in the total system, usually including the environment and the interaction between the environment and the gravitation system. We can describe the total system by the equations $-i[\mathcal{\hat{H}}_{\mu,tot}\,,\rho_{tot}]=0$. These equations are the requirements of the general covariance. Similarly to the case of the scalar field, the gravitation system as a subsystem should be treated as an open system. One can describe the open quantum gravitation system by the following equations:
\begin{equation}
\label{eq:2.34}
\frac{\delta \rho}{\delta Y^{\mu}(x)}=-i\mathrm{Tr}_{B}[\mathcal{\hat{H}}_{\mu,tot}\,,\rho_{tot}],
\end{equation}
The trace is taken over the environment. For the open gravitation system, in general one has $\delta \rho/\delta Y^{\mu}(x)\neq 0$. Equations \eqref{eq:2.34} describe the evolution of the open quantum gravitation system along the bubble time $Y^{\mu}(x)$.

Equations \eqref{eq:2.34} are the foundations of the parameterized theory for the open quantum gravitation system. The equation \eqref{eq:2.34} has a subtle difference from that of the usual von Neumann equation \eqref{eq:1.1}. The density matrix in equation \eqref{eq:1.1} evolves along the hypersurface $t=constant$, but the density matrix in equations \eqref{eq:2.34} evolves along any family of hypersurface $\sum_{t}$. Different ways of the 3+1 decomposition can lead to different sets of hypersurface $\sum_{t}$.

Equations \eqref{eq:2.34} in fact include infinitely many equations. Usually, these equations are complicated. For convenience, let us introduce the smeared version of the Hamiltonian constraint and the diffeomorphism constraints~\cite{DL,AP}:
\begin{equation}
\label{eq:2.35}
H_{tot}=\int dx^{3}N^{\mu}\mathcal{H}_{\mu,tot}=0.
\end{equation}
Here, $N^{\mu}=\dot{Y}^{\mu}$ are the Lagrangian multipliers. $N^{\mu}$ represent certain non-physical gauge degree of freedom. Different $N^{\mu}$ means different 3+1 decomposition. In equation \eqref{eq:2.35}, $H_{tot}$ is also called the Dirac Hamiltonian. One often chooses different $N^{\mu}$ according to the specific situation.

From equation \eqref{eq:2.35}, we learn that $H_{tot}$ is not dependent on $x^{a}$. Therefore, the smeared version of \eqref{eq:2.34} should be
\begin{equation}
\label{eq:2.37}
\frac{\partial \rho}{\partial t}=-i\mathrm{Tr}_{B}[{\hat{H}}_{tot}\,,\rho_{tot}].
\end{equation}
There are the interactions between the environment and the gravitation system. The entanglement entropy of the gravitational subsystem will then change with time. This means that the diagonal elements of the density matrix of the reduced gravitation system can change with time. Then,  we expect that the time derivative of the density matrix in equation \eqref{eq:2.37} are not equal to zero. We pointed out that equation \eqref{eq:2.37} can be strictly derived by introducing the Brown-Kucha$\check{\mathrm{r}}$  dust field~\cite{JD,VT,HWJ}. And the coordinate time variable $``t"$ can be reviewed as the dust field~\cite{JD,VT}.

Equations \eqref{eq:2.34} and \eqref{eq:2.37} can be used to describe the evolution of the open quantum gravitation system. Both of these two equations can describe certain interesting features of the open quantum gravitation system, such as the decoherence,  the variation of the entanglement entropy, the nonequilibrium state... If one just cares about the evolution along the time coordinate $t$, one can use the equation \eqref{eq:2.37} to describe the open quantum gravitation system.

In loop quantum gravity, the basic canonical variable are the Ashtekar variables and the densitized triad. In addition to the Hamiltonian constraint and the diffeomorphism constrains, there is also the Gaussian constraint~\cite{AJ}. The smeared version of the Gaussian constraint is the generators of the SU(2) transformations~\cite{AP,PS}.  The definition of the Dirac Hamiltonian should also contain the Gaussian constraint. For the homogenous and isotropic situation, the diffeomorphism constrains and the Gaussian constraint are trivial.

In this work, we consider the quantum evolution along the time coordinate $t$. We will use equation \eqref{eq:2.37} to describe the open quantum gravitation system. We consider a specific but important case, i.e., the environment is a large heat bath. The degree of freedom of the bath is very large. This usually leads the equation \eqref{eq:2.37} to be very difficult to solve analytically. One needs to introduce certain approximations to simplify the equation \eqref{eq:2.37} according to the specific situation.

Suppose that the Supper-Hamiltonian vector operator $\mathcal{\hat{H}}_{\mu,tot}$ of the total system can be written as the form
\begin{equation}
\label{eq:3.1}
\mathcal{\hat{H}}_{\mu,tot}=\mathcal{\hat{H}}_{\mu,grav}+\mathcal{\hat{H}}_{\mu,B}+\mathcal{\hat{H}}_{\mu,int},
\end{equation}
then the Dirac Hamiltonian (in the later, we just call it Hamiltonian) can be written as
\begin{equation}
\label{eq:3.2}
\hat{H}_{tot}=\int dx^{3} N^{\mu}\mathcal{\hat{H}}_{\mu,tot}=\hat{H}_{grav}+\hat{H}_{B}+\hat{H}_{int}.
\end{equation}
Here, $\hat{H}_{grav}$ represents the Hamiltonian operator of the pure gravity, $\hat{H}_{B}$ represents the Hamiltonian operator of the bath and $\hat{H}_{int}$ represents the interaction Hamiltonian operator between the gravitation system and the bath. Similarly, in loop quantum gravity, the Hamiltonian $H_{tot}$ should also contain the Gaussian constraint. We consider the case where the interaction Hamiltonian operator $\hat{H}_{int}$ is small compared to $\hat{H}_{B}$ and $\hat{H}_{grav}$. Thus the average of the operator $\hat{H}_{grav}$ plus the average of the operator $\hat{H}_{B}$ are approximately equal to zero.

For convenience, we transform the equation \eqref{eq:2.37} into the interaction picture:
\begin{equation}
\label{eq:3.6}
\frac{\partial \tilde{\rho}}{\partial t}=-i\mathrm{Tr}_{B}[{\tilde{H}}_{int}\,,\tilde{\rho}_{tot}].
\end{equation}
Here, $\tilde{\rho}$, $\tilde{H}_{int}$ and $\tilde{\rho}_{tot}$ represent the reduced density matrix, the interaction Hamiltonian and the total density matrix in the interaction picture, respectively.
Using the Born-Markov approximation, equation \eqref{eq:3.6} becomes~\cite{HF,HC}
\begin{equation}
\label{eq:3.7}
\frac{\partial \tilde{\rho}(t)}{\partial t}=-\int _{0}^{\infty} ds \,\mathrm{Tr}_{B}\Big\{[\tilde{H}_{int}(t),[\tilde{H}_{int}(t-s),\tilde{\rho}(t)R_{0}]]\Big\}.
\end{equation}

Equation \eqref{eq:3.7} can be used to describe the evolution of the open quantum gravitation system along the coordinate time $t$. We should point out that in equation \eqref{eq:3.7}, the interaction Hamiltonian operator has some gauge degrees of freedom (see \eqref{eq:3.1} and \eqref{eq:3.2}, $\tilde{H}_{int}=\int dx^{3} N^{\mu}\hat{H}_{\mu,int}$), and then the density matrix $\tilde{\rho}$ also has some gauge degrees of freedom. We can fixed these non-physical degrees of freedom in different ways according to the specific situations.

There exist certain systems where the total Hamiltonian can not be written in the form \eqref{eq:3.2}, equation \eqref{eq:3.7} then can not be used to describe these systems. In addition, there are certain systems where the total Hamiltonian can be written in the form \eqref{eq:3.2}. However, the density matrix of the environment can change with time. For these systems, we also can not use equation \eqref{eq:3.7} to describe them. Although both of them can be important in the open quantum gravitation system, we do not consider these problems in this work. We only study a specific type of system which can be described by Equation \eqref{eq:3.7}.

\section{A specific model: loop quantum gravity coupled with a real scalar field }
\subsection{Classical Hamiltonian of the total system}
\begin{figure}[tbp]
\centering
\includegraphics[width=5cm]{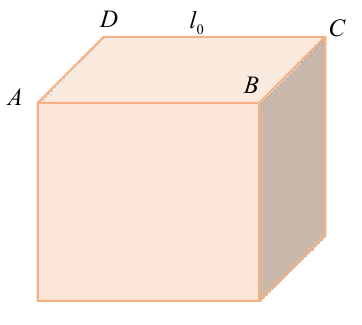}
\caption{\label{fig:1} Constraining the system to be in the cube.}
\end{figure}

For clarity, we consider a simple model to show the characteristics of the open quantum gravitation system. The system is described by the following action~\cite{LP,NB}:
\begin{eqnarray}
\label{eq:4.1}
\begin{split}
\mathcal{S}=\frac{1}{16\pi}&\int dx^{4}\sqrt{-g}R+\frac{1}{2}\int dx^{4} \sqrt{-g}g^{\mu\nu}\phi_{1,\mu}\phi_{1,\nu}
-\frac{1}{2}\int dx^{4} \sqrt{-g}m_{1}^{2}\phi_{1}^{2}\\&-\frac{1}{2} \xi\int dx^{4} \sqrt{-g}R\phi_{1}^{2}.
\end{split}
\end{eqnarray}
Here, $\phi_{1}$ represents the real scalar field, $m_{1}$ is the rest mass of the scalar field and $\xi$ is the coupling constant. The fourth term denotes the non-minimal coupling between the gravity and the scalar field. This term can be generally induced at one-loop order perturbative quantum field theory~\cite{DIE,SWM}. In this action, we do not include the Gibbons-Hawking surface term as this term does not contribute to the equation of motion. Neglecting the surface term is a common practice in quantum cosmology~\cite{AAP,AE}. In order to preserve the validity of the Born approximation,  we limit  $\xi$ as a small constant, so that the fourth term in equation \eqref{eq:4.1} is small compared with the other terms. This also leads to the action in \eqref{eq:4.1} to be approximately equivalent to the conventional Friedmann equation.
We consider a simple case where the spacetime is homogenous and isotropic:
\begin{equation}
\label{eq:4.2}
ds^{2}=-dt^{2}+a^{2}(t)(dx^{2}+dy^{2}+dz^{2}),
\end{equation}
where $a(t)$ is a scale factor. The determinant of the metric is $\sqrt{-g}=a^{3}$, and the Ricci scalar is given as~\cite{LP}
\begin{equation}
\label{eq:4.3}
R=6(\frac{\ddot{a}}{a}+\frac{\dot{a}^{2}}{a^{2}}).
\end{equation}
Dots represent the coordinate time derivative. Thus the Einstein-Hilbert action becomes
\begin{equation}
\label{eq:4.3b}
\frac{1}{16\pi}\int dx^{4}\sqrt{-g}R=-\frac{3}{8\pi}\int dx^{4}a\dot{a}^{2}+\frac{3}{8\pi}\int dx^{4}\frac{d}{dt}(a^{2}\dot{a}).
\end{equation}
The second term on the right hand side of this equation has no influence on the dynamics and can be canceled by the Gibbons–Hawking–York surface term. Then the Lagrangian of the pure gravitation is
\begin{equation}
\label{eq:4.4}
L_{grav}=-\frac{3}{8\pi}V_{0}a\dot{a}^{2},
\end{equation}
where $V_{0}\equiv\ell_{0}^{3}$ is the coordinate volume. We constrain the system to be in a cube where the coordinate length of every side is $\ell_{0}$. And the coordinate volume is $V_{0}=\ell_{0}^{3}$, see figure~\ref{fig:1}.
The conjugate momentum of the scale factor $a(t)$ is
\begin{equation}
\label{eq:4.5}
\pi_{a}=\frac{\partial L_{grav}}{\partial \dot{a}}=-\frac{3}{4\pi}V_{0}a\dot{a}\,,
\end{equation}
The Hamiltonian of the pure gravitation system is thus given as
\begin{equation}
\label{eq:4.6}
H_{grav}=\pi_{a}\dot{a}-L_{grav}=-\frac{3}{8\pi}V_{0}a\dot{a}^{2}.
\end{equation}

The Lagrangian of the real scalar field in curved space time is
 \begin{equation}
\label{eq:4.7}
L_{\phi_{1}}=\frac{1}{2}\int dx^{3} \sqrt{-g}g^{\mu\nu}\phi_{1,\mu}\phi_{1,\nu}-\frac{1}{2}\int dx^{3} \sqrt{-g}m_{1}^{2}\phi_{1}^{2}\,,
\end{equation}
From this Lagrangian, combining with the FRW metric \eqref{eq:4.2}, we can obtain the equation of motion for the scalar field as~\cite{LP}
 \begin{equation}
\label{eq:4.8}
a^{-3}\cdot\partial_{\mu}(a^{3}g^{\mu\nu}\partial_{\nu}\phi_{1})+m_{1}^{2}\phi_{1}=0.
\end{equation}
Obviously, when $a=1$, equation \eqref{eq:4.8} reduces to the ordinary Klein-Gordon equation in the Minkowski spacetime. In equation \eqref{eq:4.8}, there is a friction term ( or Hubble drag ) which is proportional to $\dot{a}$.

For simplicity, we consider the simple case that  $\dot{a}$ is small. That is, we only consider the case where the evolution of the gravitation system is very slow. This toy model may not be able to describe the early universe, but still can reveal some interesting features ( see book~\cite{LP} for more details). In this case, both the Hubble drag in \eqref{eq:4.8} and the cosmological particle production are small, thus we can neglect them~\cite{LP}. This means we approximately treat the evolution of the scalar field as adiabatic. Thus we can introduce the adiabatic approximation, expanding the scalar field as~\cite{LP}
 \begin{equation}
\label{eq:4.9}
\phi_{1}(x)=\sum_{\vec{k}}\{A_{\vec{k}}f_{\vec{k}}(x)+A_{\vec{k}}^{\dag}f_{\vec{k}}^{\ast}(x)\},
\end{equation}
where
\begin{equation}
\label{eq:4.10}
f_{\vec{k}}(x)=\frac{1}{\sqrt{V_{0}a^{3}}}\frac{1}{\sqrt{2\omega_{k}}}exp\{i(\vec{k}\cdot\vec{x}-\int_{0}^{t}\omega_{k}dt)\},
\end{equation}
where $\omega_{k}=(\vec{k}^{2}/a^{2}+m_{1}^{2})^{1/2}$. Under the adiabatic approximation, \eqref{eq:4.9} is the general solution of the equation \eqref{eq:4.8}.

The Hamiltonian of the scalar field is given as
\begin{equation}
\label{eq:4.15}
H_{\phi_{1}}=-\frac{1}{2}\int dx^{3} \sqrt{-g}\left(\dot{\phi_{1}}^{2}-g^{aa}\phi_{1}\partial_{a}^{2}\phi_{1}-m_{1}^{2}\phi_{1}^{2}\right).
\end{equation}
Combine \eqref{eq:4.15} and the equation of motion \eqref{eq:4.8}, we have
\begin{equation}
\label{eq:4.16}
H_{\phi_{1}}\approx-\frac{1}{2}\int dx^{3} \sqrt{-g}\left(\dot{\phi_{1}}^{2}-\phi_{1}\partial_{0}^{2}\phi_{1}\right)
\end{equation}
Here, as $\dot{a}$ is small, thus we neglected all the terms including $\dot{a}$. Substituting the equations \eqref{eq:4.9} and \eqref{eq:4.10} into \eqref{eq:4.16}, one can obtain
\begin{equation}
\label{eq:4.17}
H_{\phi_{1}}\approx\sum_{\vec{k}}\omega_{k}A_{\vec{k}}^{\dag}A_{\vec{k}}.
\end{equation}
Here, $\omega_{k}=(\vec{k}^{2}/a^{2}+m_{1}^{2})^{1/2}$. Equation \eqref{eq:4.17} is consistent with the results in ~\cite{ETA,SSF}. Again, in \eqref{eq:4.17}, we neglected all the terms including $\dot{a}$.  All of these terms are small compared with $\sum_{\vec{k}}\omega_{k}A_{\vec{k}}^{\dag}A_{\vec{k}}$. But some of these terms maybe not very small compared with the non-minimal coupling interaction Hamiltonian. For simplicity, we do not consider this case in this work. When $\dot{a}$ is small, equation \eqref{eq:4.17} is approximately correct in the curved spacetime while it is strictly right in the Minkowski spacetime.

The interaction Hamiltonian is given as
\begin{equation}
\label{eq:4.18}
H_{int}=\frac{1}{2}\xi\int dx^{3}\sqrt{-g}R\phi_{1}^{2}=3\xi a^{3}(\frac{\ddot{a}}{a}+\frac{\dot{a}^{2}}{a^{2}})\int dx^{3} \phi_{1}^{2}.
\end{equation}
As we assume that $\xi$ is a small constant, we can use the following Friedmann equation to describe the classical evolution of the space~\cite{SD}
\begin{equation}
\label{eq:4.19}
(\frac{\dot{a}}{a})^2=\frac{8\pi}{3}\rho_{\phi_{1}}+o(\xi),
\end{equation}
\begin{equation}
\label{eq:4.20}
\frac{\ddot{a}}{a}=-\frac{4\pi}{3}(\rho_{\phi_{1}} +3P_{\phi_{1}})+o(\xi).
\end{equation}
 Here, $o(\xi)$ represents the first order small quantity of $\xi$. Assume that the mass $m_{1}$ is big enough, we approximately treat the scalar particles as the non-relativistic particles, then $P_{\phi_{1}}=0$~\cite{SD}.  Combine with the Friedmann equations, we have $\dot{a}^{2}/a^{2}=-2\ddot{a}/a$. Thus, the interaction Hamiltonian \eqref{eq:4.18} can be written as
\begin{equation}
\label{eq:4.21}
H_{int}=-3\xi a^{2}\ddot{a}\int dx^{3} \phi_{1}^{2}+o(\xi^{2}).
\end{equation}
Combining with \eqref{eq:4.9} and \eqref{eq:4.10}, we have
\begin{eqnarray}\begin{split}
\label{eq:4.22}
H_{int}= -\frac{3\xi (2\pi)^{3}}{2V_{0}}&\cdot\frac{\ddot{a}}{a}\sum_{\vec{k}}\frac{1}{\omega_{k}}\Big\{A_{\vec{k}}A_{-\vec{k}}\,exp(-2i\int_{0}^{t}\omega_{k}dt)\\
                                                                                                             &+A_{\vec{k}}^{\dag}A_{-\vec{k}}^{\dag}\,exp(2i\int_{0}^{t}\omega_{k}dt)+A_{\vec{k}}^{\dag}A_{\vec{k}}+A_{\vec{k}}A_{\vec{k}}^{\dag}\Big\}+o(\xi^{2}).
\end{split}
\end{eqnarray}
Taking the continuous limit and replacing $\sum_{\vec{k}}$ by $(2\pi)^{-3}V_{0}\int d\vec{k}^{3}$.
Noting $\int d\vec{k}^{3}= 4\pi a^{3}\int \omega (\omega^{2}-m_{1}^{2})^{1/2}d\omega$, and carrying out the partial integration, then \eqref{eq:4.22} approximately becomes ( The boundary term induced by the partial integration can be eliminated by the Gibbons-Hawking surface term.)
\begin{eqnarray}\begin{split}
\label{eq:4.24}
H_{int}\approx12i\pi&\xi\dot{a}\cdot a^{2}\int d\omega\cdot\omega\sqrt{\omega^{2}-m_{1}^{2}}\Big\{A_{\vec{k}}^{\dag}A_{-\vec{k}}^{\dag}\,exp(2i\int_{0}^{t}\omega_{k}dt)\\
    &-A_{\vec{k}}A_{-\vec{k}}\,exp(-2i\int_{0}^{t}\omega_{k}dt)\Big\}.
\end{split}
\end{eqnarray}
The integral interval of the variable $\omega$ is from $m_{1}$ to infinity. As $\dot{a}$ is small, the volume of the cube slowly changes with time. We can then approximately set
\begin{equation}
\label{eq:4.25}
V(t)= V(0)+o(\dot{a}),
\end{equation}
where $V(t)$ is the physical volume of the cube at the moment $t$ ,  $V(0)$ is the physical volume of the cube at the initial time, and $o(\dot{a})$ is a small quantity ( $V(0)$ should be distinguished with $V_{0}$, $V(0)$ represents the physical volume at the initial time, $V_{0}$ represents the coordinate volume ). Bring \eqref{eq:4.25} into \eqref{eq:4.24}, we have
\begin{eqnarray}\begin{split}
\label{eq:4.26}
H_{int}=\frac{12i\pi\xi}{\ell_{0}^{3}}&V(0)\cdot\frac{\dot{a}}{a}\int d\omega\cdot\omega\sqrt{\omega^{2}-m_{1}^{2}}\Big\{A_{\vec{k}}^{\dag}A_{-\vec{k}}^{\dag}\,exp(2i\int_{0}^{t}\omega_{k}dt)\\
    &-A_{\vec{k}}A_{-\vec{k}}\,exp(-2i\int_{0}^{t}\omega_{k}dt)\Big\}+o(\xi\dot{a}^2).
\end{split}
\end{eqnarray}

To sum up, the Hamiltonian of the total system can be written as
\begin{equation}
\label{eq:4.27}
H_{tot}=H_{grav}+H_{\phi_{1}}+H_{int},
\end{equation}
where,
\begin{eqnarray}\begin{split}
\label{eq:4.28}
&H_{grav}=-\frac{3}{8\pi}V_{0}a\dot{a}^{2},\\
&H_{\phi_{1}}\approx\sum_{\vec{k}}\omega_{k}A_{\vec{k}}^{\dag}A_{\vec{k}},\\
&H_{int}=\frac{12i\pi\xi}{\ell_{0}^{3}}V(0)\cdot\frac{\dot{a}}{a}\int d\omega\cdot\omega\sqrt{\omega^{2}-m_{1}^{2}}\Big\{A_{\vec{k}}^{\dag}A_{-\vec{k}}^{\dag}\,exp(2i\int_{0}^{t}\omega_{k}dt)\\
    &\qquad\qquad-A_{\vec{k}}A_{-\vec{k}}\,exp(-2i\int_{0}^{t}\omega_{k}dt)\Big\}.
\end{split}
\end{eqnarray}
The Hamiltonian $H_{tot}$ determines the classical dynamics of the total system. We should point out that in the Hamiltonian $H_{\phi_{1}}$, there is a contribution from the minimal coupling between the scalar field and the gravitation system. We have not explicitly included this minimal coupling into the interaction Hamiltonian $H_{int}$. This will not lead to the logical contradiction.  From \eqref{eq:4.17}, we learn that after we quantize this system, the main influence of the minimal coupling to the scalar field is to change the frequency of the scalar particle. If we assume that the scalar field is in the thermal equilibrium state and treat the scalar field as a large bath, the main influence of the minimal coupling is to change the temperature of the bath. As we constrain $\dot{a}$ to be small, in an enough long time period (longer than the relaxation time scale of the system), one can approximately think of the temperature of the heat bath as not being changed with time.

Equations \eqref{eq:4.27} and \eqref{eq:4.28} have fully defined our total system. This is a typical open system problem. After carrying out the procedure of the quantization, we can study it by the quantum master equation \eqref{eq:3.7}. Usually, if the subsystem just interacts with one heat bath, the subsystem will reach an equilibrium state at longtimes. Thus, we expect that after the transient relaxation, the gravitational system will reach the equilibrium state. We will prove this point in the next.

\subsection{Quantization for the total system}

\subsubsection{Quantization for the gravity}
In our model, the evolution of the space is driven by the scalar particles. The classical evolution of the space is described by the Friedmann equations. From the Friedmann equations \eqref{eq:4.19} and \eqref{eq:4.20}, we can obtain $a\propto t^{2/3}$ and $\dot{a}\propto t^{-1/3}$. Therefore, in quantum cosmology, when the universe expands to be larger, we expect that the average value of the scale factor $a$ should be approximately proportional to the $2/3$ power of the coordinate time, i.e. $\langle a\rangle\propto t^{2/3}$. The average value of the variation rate of the scale expansion factor $a$ should be approximately proportional to the inverse of the $1/3$ power of the coordinate time, i.e., $\langle\dot{a}\rangle\propto t^{-1/3}$. Thus, this period when $\dot{a}$ is small may not be at the extremely early times of the universe. Nevertheless, as a toy model, we quantize the spacetime by the way of the loop quantum gravity. This can simplify our calculations.

In loop quantum gravity, the basic canonical variables are the Ashtekar variable ($A_{a}^{i}$) and its conjugate momentum ($E^{a}_{i}$). They are defined as~\cite{AP}
\begin{equation}
\label{eq:4.29}
A_{a}^{i}=-\frac{1}{2}\omega_{ak}^{j}\epsilon^{ik}_{j}+\gamma K_{a}^{i},
\end{equation}
\begin{equation}
\label{eq:4.30}
E^{a}_{i}=det(e^{a}_{i})\cdot e^{a}_{i}.
\end{equation}
Here, $a,\,i, \,j, \,k=1, \,2, \,3$, $\omega^{j}_{ak}$ is the spin connection and $K_{a}^{i}$ is the extrinsic curvature. $\gamma$ is the Barbero-Immirzi parameter. $e^{a}_{i}$ is the triad and $det(e^{a}_{i})$ represents the determinant of the triad. $\epsilon^{ik}_{j}$ is the signature of permutation of (123). For more detailed explanation about these quantities, see~\cite{CK,CR,AP,AJ,PS}.
Considering the thermodynamics of the black hole, the Barbero-Immirzi parameter is usually fixed as $\gamma=\mathrm{ln}2/(\pi\sqrt{3})$~\cite{CR,AJA}.

When the spacetime is described by the FRW metric \eqref{eq:4.2}, from \eqref{eq:4.29} and \eqref{eq:4.30}, we can obtain the Ashtekar variables $A_{a}^{i}$ and the associated conjugate momentum $E^{a}_{i}$ which are $A_{a}^{i}=diag(\gamma\dot{a},\gamma\dot{a},\gamma\dot{a})$ and $E^{a}_{i}=diag(a^{2},a^{2},a^{2})$, respectively~\cite{AAP}. Thus we can treat $\gamma\dot{a}$ and $a^{2}$ as the basic canonical variables for the homogenous and isotropic spacetime. The Poisson bracket between these two variables are~\cite{AAP}
\begin{equation}
\label{eq:4.33}
\{\gamma\dot{a}, a^{2}\}=\frac{8\pi\gamma}{3V_{0}}.
\end{equation}

It is more convenient to use the following canonical variables instead of the Ashtekar variables $A_{a}^{i}$ and $E^{a}_{i}$~\cite{AAP,AMJ,DC,DCC}:
\begin{equation}
\label{eq:4.34}
c=\gamma \ell_{0} \dot{a}\,;
\end{equation}
\begin{equation}
\label{eq:4.35}
p=\ell_{0}^{2}a^{2}.
\end{equation}
From the definition of the variables $c$ and $p$, we learn that $c$ is proportional to $\dot{a}$, so it describes the variation rate of the space geometry. $c/\ell_{0}$ is the diagonal element of the Ashtekar-Barbero variables. $|p|$ is the physical area of the square on the surface of the cube (such as the square $ABCD$ in figure~\ref{fig:1}). It has a simple relationship with the volume of the cube: $V=|p|^{\frac{3}{2}}$.

The Poisson bracket of the variable $c$ and its conjugate momentum $p$ is~\cite{AAP}
\begin{equation}
\label{eq:4.37}
\{c,p\}=\frac{8\pi\gamma}{3}.
\end{equation}
Comparing \eqref{eq:4.37} and \eqref{eq:4.33}, we found  that the Poisson bracket \eqref{eq:4.37} does not depend on the coordinate volume $V_{0}$. This is the main reason why we introduce the variables $c$ and $p$. Bringing \eqref{eq:4.34} and \eqref{eq:4.35} into \eqref{eq:4.6}, the Hamiltonian of the pure gravitation becomes
\begin{equation}
\label{eq:4.38}
H_{grav}=\frac{-3}{8\pi\gamma^{2}}c^{2}\sqrt{|p|}\,.
\end{equation}
Noted that $p$ represents the physical area of the square, thus $p$ can not be negative. Therefore equation \eqref{eq:4.38} is well-defined.

In loop quantum gravity, the eigenvalue of the area operator $\hat{A}_{s}$ is discrete~\cite{CR},
\begin{equation}
\label{eq:4.39}
\hat{A}_{s}|S\rangle=8\pi\gamma\sum_{p}\sqrt{j_{p}(j_{p}+1)}\,|S\rangle\,;\quad j_{p}=0,\frac{1}{2},1,\frac{3}{2},2,...
\end{equation}
Here, $|S\rangle$ is the eigenvector of the area operator $\hat{A}_{s}$, the eigenvalue of $\hat{A}_{s}$ represents the area of the surface. The summation is taken over all the paths  in the state $|S\rangle$ which cross the surface. All the eigenvalues of the operator $\hat{A}_{s}$ form a discrete spectrum. The smallest element of the area is $\triangle=4\pi\gamma\sqrt{3}$~\cite{CR}.

It is natural to think that the square $ABCD$ in figure~\ref{fig:1} is composed of some smallest elements. The smallest element acts as a very small square, making the coordinate length of the edge of the smallest element as $\bar{\mu}\ell_{0}$, where $\bar{\mu}$ is called the discreteness parameter. Assume that the square $ABCD$ contains $M$ smallest elements, then we have~\cite{AAP,AE}
\begin{equation}
\label{eq:4.41}
M\triangle=|p|,
\end{equation}
\begin{equation}
\label{eq:4.42}
M(\bar{\mu}\ell_{0})^{2}=\ell_{0}^{2}.
\end{equation}
\eqref{eq:4.41} means that the physical area of the square $ABCD$ is equal to the total physical area of all smallest elements. \eqref{eq:4.42} means the coordinate area of the square $ABCD$ is equal to the total coordinate area of all the smallest elements. Combining \eqref{eq:4.41} and \eqref{eq:4.42}, we can express the discreteness parameter $\bar{\mu}$ as~\cite{AAP,AE}
\begin{equation}
\label{eq:4.43}
\bar{\mu}=\sqrt{\frac{\triangle}{|p|}}\,.
\end{equation}

In order to quantize the gravity, one should do the following replacement:
\begin{equation}
\label{eq:4.44}
c\rightarrow\hat{c}=i\frac{8\pi\gamma}{3}\cdot\frac{d}{dp};
\end{equation}
\begin{equation}
\label{eq:4.45}
p\rightarrow\hat{p}=p;
\end{equation}
Hence the Hamiltonian operator of the pure gravitation is given as
\begin{equation}
\label{eq:4.47}
\hat{H}_{grav}=\frac{-3}{16\pi\gamma^{2}}\big(\hat{c}^{2}\sqrt{|\hat{p}|}+\sqrt{|\hat{p}|}\hat{c}^{2}\big).
\end{equation}
The definition of the Hamiltonian operator for the quantum system is frequently plagued by the ordering ambiguities. In \eqref{eq:4.47}, we choose the symmetrized factor ordering so that the Hamiltonian operator is  Hermitian. The Hermiticity of the Hamiltonian operator can not fully fix the factor ordering. For example, $-3\hat{c}\sqrt{|\hat{p}|}\hat{c}/(8\pi\gamma^{2})$  is also a Hermitian operator and is different from the operator \eqref{eq:4.47}. The factor ordering used in \eqref{eq:4.47} is different from that in ~\cite{AAP,AMJ}. However, from equation \eqref{eq:A5} to \eqref{eq:A6}, we neglected the higher order effect of $\hbar$ (see appendix~\ref{sec:A} ). Thus different factor ordering gives rise to the same results. Generally, If we neglect the higher order effect of $\hbar$, the ordering ambiguities can be ignored. Please see~\cite{EAJ} for more details discussion about this point.

The kinematical Hilbert space of loop quantum cosmology is spanned by the set of functions $\{exp(i\mu c/2);\ \mu\in \mathbb{R}\}$ with the inner product $\langle exp(i\mu_{1}/2)|exp(i\mu_{2}/2)\rangle=\delta_{\mu_{1},\mu_{2}}$. For more details discussions about the definition of the kinematical Hilbert space, see references~\cite{AAP,AMJ,JBEL,NBDF}. Since the basic variable of the state function in the kinematic Hilbert space of the loop quantum gravity is the holonomy, rather than the variable $c$, one can not use \eqref{eq:4.47} straightforwardly. One needs to do the following replacement~\cite{AAP,DC}:
\begin{equation}
\label{eq:4.48}
\hat{c}\rightarrow \frac{sin(\bar{\mu}\hat{c})}{\bar{\mu}}.
\end{equation}
Consequently, the Hamiltonian operator in \eqref{eq:4.47} should be replaced by
\begin{equation}
\label{eq:4.49}
\hat{H}_{grav}=\frac{-3}{16\pi\gamma^{2}}\Big\{\frac{sin^{2}(\bar{\mu}\hat{c})}{\bar{\mu}^{2}}\sqrt{|\hat{p}|}+\sqrt{|\hat{p}|}\frac{sin^{2}(\bar{\mu}\hat{c})}{\bar{\mu}^{2}}\Big\}.
\end{equation}
When the discreteness parameter $\bar{\mu}$ approaches zero, the Hamiltonian in \eqref{eq:4.49} tends to become \eqref{eq:4.47}. Thus, we expect that when the discreteness parameter $\bar{\mu}$ approaches zero, the loop quantum cosmology becomes the usual canonical quantum cosmology.

In the following, we will neglect the operator hat on the variables $c$ and $p$. Combine \eqref{eq:4.43} and \eqref{eq:4.49}, then we have
\begin{equation}
\label{eq:4.50}
\hat{H}_{grav}=\frac{3}{64\gamma^{2}\triangle}\Big\{\Big(e^{2i\bar{\mu}c}-2+e^{-2i\bar{\mu}c}\Big)|p|^{\frac{3}{2}}+|p|^{\frac{3}{2}}\Big(e^{2i\bar{\mu}c}-2+e^{-2i\bar{\mu}c}\Big)\Big\}.
\end{equation}
For convenience, define~\cite{AE}
\begin{equation}
\label{eq:4.51}
\mathcal{V}\equiv\frac{|p|^{\frac{3}{2}}}{2\pi\gamma\sqrt{\triangle}}sgn(p),
\end{equation}
$\mathcal{V}$ is a dimensionless operator\footnote{ Sometimes we neglect the operator hat. Readers can easily identify what is the $c$-number and what is the $q$-number according to the related contents.}. $sgn(p)$ is the sign of $p$. $sgn(p)=1$ if $p>0$ and $sgn(p)=-1$ if $p<0$~\cite{AAP}. Certain relations emerge~\cite{AAP,AE}:
\begin{equation}
\label{eq:4.55}
|p|^{\frac{3}{2}} \Psi(\mathcal{V})=2\pi\gamma \sqrt{\triangle}\mathcal{|V|}\Psi(\mathcal{V});
\end{equation}
\begin{equation}
\label{eq:4.56}
exp(\pm i\bar{\mu}c)|\mathcal{V}_{0}\rangle=exp(\mp 2\frac{d}{d\mathcal{V}})|\mathcal{V}_{0}\rangle=|\mathcal{V}_{0}\pm2\rangle.
\end{equation}
Here, $\Psi(\mathcal{V})$ is any function of the dimensionless quantity $\mathcal{V}$, and $|\mathcal{V}_{0}\rangle=\delta(\mathcal{V}-\mathcal{V}_{0})$ is the eigenvector of the operator $\hat{\mathcal{V}}$. $|\mathcal{V}_{0}\rangle$ represents the state where the volume is $\mathcal{V}_{0}$,
\begin{equation}
\label{eq:4.57}
\hat{\mathcal{V}}|\mathcal{V}_{0}\rangle=\mathcal{V}_{0}|\mathcal{V}_{0}\rangle.
\end{equation}

Defining
\begin{equation}
\label{eq:4.58}
\sigma^{\pm}\equiv exp(\pm i\bar{\mu}c),
\end{equation}
we can prove the following relations:
\begin{equation}
\label{eq:4.59}
[\sigma^{+},\,\sigma^{-}]=0;
\end{equation}
\begin{equation}
\label{eq:4.60}
[\sigma^{\pm},\,\mathcal{V}]=\mp2\sigma^{\pm};
\end{equation}
\begin{equation}
\label{eq:4.61}
\sigma^{\pm}|\mathcal{V}_{0}\rangle=|\mathcal{V}_{0}\pm2\rangle;
\end{equation}
\begin{equation}
\label{eq:4.62}
\sigma^{\pm}\Psi(\mathcal{V})=\Psi(\mathcal{V}\mp2).
\end{equation}
\eqref{eq:4.59} and \eqref{eq:4.60} show that $\sigma^{+}$  commutes with $\sigma^{-}$, $\sigma^{\pm}$ does not commute with the dimensionless volume operator $\mathcal{V}$. \eqref{eq:4.61} shows that $\sigma^{+}$ and $\sigma^{-}$ can be treated as the raising and lowering operators of the volume, respectively. If we use $\sigma^{\pm}$ to act on a state where the volume is $\mathcal{V}_{0}$, then we reach a state with the volume $\mathcal{V}_{0}\pm2$.

Combining \eqref{eq:4.50}, \eqref{eq:4.51} and \eqref{eq:4.58}, the Hamiltonian operator in \eqref{eq:4.50} can be written as
\begin{equation}
\label{eq:4.63}
\hat{H}_{grav}=\frac{3}{32\gamma\sqrt{\triangle}}\Big\{\Big((\sigma^{+})^{2}+(\sigma^{-})^{2}-2\Big)\mathcal{|V|}+\mathcal{|V|}\Big((\sigma^{+})^{2}+(\sigma^{-})^{2}-2\Big)\Big\}.
\end{equation}
Setting $\mathcal{V}_{0}=2n$, where $n=1,2,3...$
Then we have
\begin{equation}
\label{eq:4.65}
\mathcal{V}|n\rangle=2n|n\rangle,
\end{equation}
\begin{equation}
\label{eq:4.66}
\sigma^{\pm}|n\rangle=|n\pm1\rangle.
\end{equation}
\eqref{eq:4.65} shows that the state $|n\rangle$ represents a state where the magnitude of the volume is $2n$. The quantity $n$ represents the serial number of the quantum state. $n=1$ means the first state with the volume $\mathcal{V}=2$, $n=2$ means the second state with the volume  $\mathcal{V}=4$,... Compare with the concept of field quanta, we can also think of the state $|n\rangle$ as the one including $n$ space quanta, and the volume of each space quanta as being $\mathcal{V}=2$. In this way, one can think the macroscopically space as composed of very small space quanta. This is similar to the concept of the classical electromagnetic field as composed by many photons.
Noted that $\mathcal{V}_{0}\neq 0$, the reason is that in loop quantum cosmology, the universe can not switch from the zero volume state to the non-zero volume state~\cite{ATP,ATPP,AATP}.
In addition, according to equation \eqref{eq:4.51}, the value of $\mathcal{V}$ can be negative. However, in the absence of fermions, the map $\mathcal{V}\rightarrow -\mathcal{V}$ corresponds to large gauge transformations and do not change physics. Thus in loop quantum cosmology, it is customary to work with the symmetric representation, that is $\Psi(\mathcal{V})=\Psi(\mathcal{-V})$~\cite{AAP}. Thus $\Psi(\mathcal{V})$ and $\Psi(\mathcal{-V})$ corresponds to the same physics. Therefore, in this work, we do not consider these states which  $\mathcal{V}$ is negative. And the physical Hilbert space is spanned by the set of vectors $\{|n\rangle; n=1,2,3,...\}$.

\subsubsection{Quantization for the scalar field}
It is difficult to quantize the field theory in the curved spacetime. One of the reasons is that the vacuum state of the field in curved spacetime is not unique~\cite{RW}. Usually, the annihilation operator and the creation operator are different in different times, this leads to the cosmological particle production~\cite{U}. The influence of the cosmological particle production to the spacetime structure is very small compared to that of the scalar field, so we neglect this effect here and approximately treat the vacuum of the quantum scalar field as unique. In  addition, although the vacuum energy of the scalar field can also impact the spacetime structure, for simplicity, we also do not consider this effect. Above all, the Hamiltonian operator of the scalar field can be approximated as
\begin{equation}
\label{eq:4.68}
\hat{H}_{\phi_{1}}=\sum_{\vec{k}}\omega_{k}\hat{A}_{\vec{k}}^{\dag}\hat{A}_{\vec{k}}.
\end{equation}
Here, $\hat{A}^{\dag}_{\vec{k}}$ and $\hat{A}_{\vec{k}}$ are the creation and annihilation operators of the scalar particles, respectively. Noted that in \eqref{eq:4.68}, $\omega_{k}=(\vec{k}^{2}/a^{2}+m_{1}^{2})^{1/2}$, after we quantized the gravity, the scale factor $a$ is an operator.  However, as we assumed that $\dot{a}$ is small, thus we approximately treat the temperature of the bath as being unchanged in the time interval of the system relaxation. That is, in this time interval, the influence of the spacetime to the bath can be neglected. Microscopically, the influence of the spacetime to the frequency of the particles of the bath can be neglected. Therefore, we still can approximately treat the frequency as a $c$-number.

\subsubsection{Quantization for the interaction Hamiltonian}
Bringing \eqref{eq:4.34} and \eqref{eq:4.35} into \eqref{eq:4.26}, the classical interaction Hamiltonian can be written as
\begin{eqnarray}\begin{split}
\label{eq:4.69}
H_{int}=\frac{12i\pi\xi}{\gamma\ell_{0}^{3}}&V(0)\cdot\frac{c}{\sqrt{|p|}}\int d\omega\cdot\omega\sqrt{\omega^{2}-m_{1}^{2}}\Big\{A_{\vec{k}}^{\dag}A_{-\vec{k}}^{\dag}\,exp(2i\int_{0}^{t}\omega_{k}dt)\\
    &-A_{\vec{k}}A_{-\vec{k}}\,exp(-2i\int_{0}^{t}\omega_{k}dt)\Big\}.
\end{split}
\end{eqnarray}
Combining with \eqref{eq:4.48} and \eqref{eq:4.58}, and replacing the $c$-number by the corresponding $q$-number, then we obtain the interaction Hamiltonian operator
\begin{eqnarray}\begin{split}
\label{eq:4.70}
\hat{H}_{int}= &\int d\omega\cdot\omega\sqrt{\omega^{2}-m_{1}^{2}}\Big\{A_{\vec{k}}^{\dag}A_{-\vec{k}}^{\dag}\,exp(2i\int_{0}^{t}\omega_{k}dt) -A_{\vec{k}}A_{-\vec{k}}\,exp(-2i\int_{0}^{t}\omega_{k}dt)\Big\}\cdot\\
&\quad\Big(\frac{6\pi V(0)\xi}{\gamma\ell_{0}^{3}\sqrt{\triangle}}\cdot\sigma^{+}-\frac{6\pi V(0)\xi}{\gamma\ell_{0}^{3}\sqrt{\triangle}}\cdot\sigma^{-}\Big).
\end{split}
\end{eqnarray}
Note that the operators $A_{\vec{k}}^{\dag}\,exp(i\int_{0}^{t}\omega_{k}dt)$ and $A_{\vec{k}}\,exp(-i\int_{0}^{t}\omega_{k}dt)$ are in the interaction picture while the operators $\sigma^{+}$ and $\sigma^{-}$ are in the schr\"{o}dinger picture. In order to obtain a well-defined interaction Hamiltonian operator, we transform the operators $\sigma^{\pm}$ from the schr\"{o}dinger picture into the interaction picture by the following unitary transformation
\begin{equation}
\label{eq:4.71}
\sigma_{I}^{\pm}=exp\big\{i\hat{H}_{grav}t\big\}\sigma^{\pm}exp\big\{-i\hat{H}_{grav}t\big\}.
\end{equation}

The correct interaction Hamiltonian operator (in the interaction picture) is given by
\begin{equation}
\label{eq:4.72}
\hat{H}_{int}= \Gamma_{1} +\Gamma_{2},
\end{equation}
where
\begin{eqnarray}\begin{split}
\label{eq:4.73}
\Gamma_{1}\equiv&\frac{6\pi V(0)\xi}{\gamma\ell_{0}^{3}\sqrt{\triangle}}\cdot\sigma_{I}^{+}\cdot\int d\omega\cdot\omega\sqrt{\omega^{2}-m_{1}^{2}}\Big\{A_{\vec{k}}^{\dag}A_{-\vec{k}}^{\dag}\,exp(2i\int_{0}^{t}\omega_{k}dt)\\ &\qquad-A_{\vec{k}}A_{-\vec{k}}\,exp(-2i\int_{0}^{t}\omega_{k}dt)\Big\},
\end{split}
\end{eqnarray}
\begin{eqnarray}\begin{split}
\label{eq:4.74}
\Gamma_{2}\equiv&\frac{-6\pi V(0)\xi}{\gamma\ell_{0}^{3}\sqrt{\triangle}}\cdot\sigma_{I}^{-}\cdot\int d\omega\cdot\omega\sqrt{\omega^{2}-m_{1}^{2}}\Big\{A_{\vec{k}}^{\dag}A_{-\vec{k}}^{\dag}\,exp(2i\int_{0}^{t}\omega_{k}dt)\\ &\qquad-A_{\vec{k}}A_{-\vec{k}}\,exp(-2i\int_{0}^{t}\omega_{k}dt)\Big\},
\end{split}
\end{eqnarray}
Assuming that the scalar field is in the thermal equilibrium state, then
\begin{equation}
\label{eq:4.79}
\sigma_{I}^{\pm}=\sigma^{\pm}exp\big\{\pm2i \langle\mathcal{\hat{H}}\rangle t\big\}
\end{equation}
and
\begin{equation}
\label{eq:4.86}
\langle\mathcal{H}\rangle\equiv -\frac{\gamma \sqrt{\Delta}e^{-\beta_{1}m_{1}}}{\ell_{0}^{3}}\Big(128\pi^{2} m_{1}^{4}T_{1}^{3}+6\sqrt{2}\pi^{\frac{5}{2}}m_{1}^{\frac{3}{2}}T_{1}^{\frac{5}{2}}\Big).
\end{equation}
Here, $\beta_{1}=1/T_{1}$ and $T_{1}$ is the temperature of the bath. The main steps of the derivations for equations \eqref{eq:4.79} and \eqref{eq:4.86} are summarized in appendix~\ref{sec:A}.

To sum up, the total system is described by the following Hamiltonian operator:
\begin{equation}
\label{eq:4.87}
\hat{H}_{tot}=\hat{H}_{grav}+\hat{H}_{\phi_{1}}+\hat{H}_{int},
\end{equation}
where
\begin{eqnarray}\begin{split}
\label{eq:4.88}
\hat{H}_{grav}=\frac{3}{32\gamma\sqrt{\triangle}}\Big\{\Big(&(\sigma^{+})^{2}+(\sigma^{-})^{2}-2\Big)|\mathcal{V}|+|\mathcal{V}|\Big((\sigma^{+})^{2}+(\sigma^{-})^{2}-2\Big)\Big\},\\
&\hat{H}_{\phi_{1}}=\sum_{\vec{k}}\omega_{k}\hat{A}_{\vec{k}}^{\dag}\hat{A}_{\vec{k}}\,,\\
&\hat{H}_{int}= \Gamma_{1} +\Gamma_{2}\,,
\end{split}
\end{eqnarray}
\begin{figure}[tbp]
\centering
\includegraphics[width=11cm]{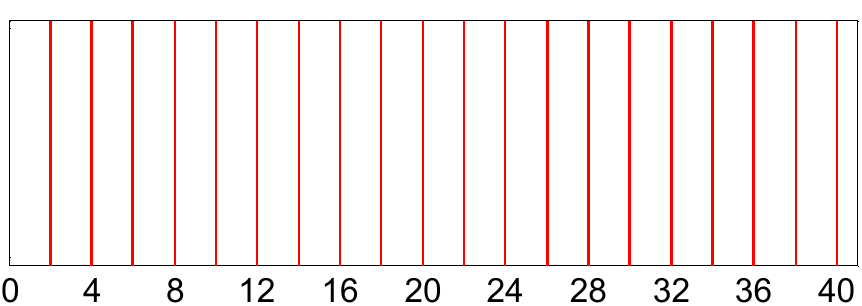}
\caption{\label{fig:2} The spectrum of the quantum state of the space.}
\end{figure}
$\Gamma_{1}$  and $\Gamma_{2}$ are defined by \eqref{eq:4.73} and \eqref{eq:4.74}, respectively.

$\Gamma_{1}$ and $\Gamma_{2}$ contain the linear terms of $\sigma^{+}$ and $\sigma^{-}$, respectively. If we use $\sigma^{\pm}$  to act on the state $|\mathcal{V}_{0}\rangle$, then this state would be changed to $|\mathcal{V}_{0}\pm2\rangle$. The magnitude of the volume which can be observed are $\mathcal{V}_{0}=2,4,6,8,...$ as shown by the spectrum composed of the red vertical lines in figure~\ref{fig:2}.
Equations \eqref{eq:4.87} and \eqref{eq:4.88} have defined a typical open quantum system.  The system has a quantum degree of freedom and coupled to a heat bath with infinite degrees of freedom. According to the quantum master equation, the system will reach the equilibrium state after the transient relaxation. We will solve the quantum master equation and obtain this equilibrium state. This is the core of the next section. All other analysis are based on this state.

\subsection{The characteristics of quasi-steady state}
\subsubsection{Quasi-steady state of quantum spacetime}
The total Hamiltonian operator \eqref{eq:4.87} determines the quantum natures of the total system. In this work, we focus on the gravitational subsystem. Its evolution can be described by the quantum master equation \eqref{eq:3.7}. Substituting \eqref{eq:4.72} into \eqref{eq:3.7}, we obtain
\begin{eqnarray}\begin{split}
\label{eq:4.89}
\dot{\rho}=-\sum_{i=1}^{2}\sum_{j=1}^{2}\int_{0}^{\infty} ds \,&\mathrm{Tr}_{B}\Big\{\Gamma_{i}(t)\Gamma_{j}(t-s)\rho(t)R_{0}-\Gamma_{i}(t)\rho(t)R_{0}\Gamma_{j}(t-s)\\
&-\Gamma_{i}(t-s)\rho(t)R_{0}\Gamma_{j}(t)+\rho(t)R_{0}\Gamma_{i}(t-s)\Gamma_{j}(t)\Big\}.
\end{split}
\end{eqnarray}
Note that in \eqref{eq:4.89}, all operators are defined in the interaction picture. Defining
\begin{eqnarray}\begin{split}
\label{eq:4.90}
&E_{ij}\equiv \int_{0}^{\infty}ds\,\mathrm{Tr}_{B}\Big\{\Gamma_{i}(t)\Gamma_{j}(t-s)\rho(t)R_{0}\Big\},\\
&F_{ij}\equiv \int_{0}^{\infty}ds\,\mathrm{Tr}_{B}\Big\{\Gamma_{i}(t)\rho(t)R_{0}\Gamma_{j}(t-s)\Big\},\\
&G_{ij}\equiv \int_{0}^{\infty}ds\,\mathrm{Tr}_{B}\Big\{\Gamma_{i}(t-s)\rho(t)R_{0}\Gamma_{j}(t)\Big\},\\
&H_{ij}\equiv \int_{0}^{\infty}ds\,\mathrm{Tr}_{B}\Big\{\rho(t)R_{0}\Gamma_{i}(t-s)\Gamma_{j}(t)\Big\},
\end{split}
\end{eqnarray}
then \eqref{eq:4.89} can be written as
\begin{equation}
\label{eq:4.91}
\dot{\rho}=-\sum_{i=1}^{2}\sum_{j=1}^{2}\Big(E_{ij}-F_{ij}-G_{ij}+H_{ij}\Big).
\end{equation}
So the evolution equation of the reduced density matrix $\rho$ in the volume representation is given as
\begin{equation}
\label{eq:4.92}
\langle n|\dot{\rho}|m\rangle=-\sum_{i=1}^{2}\sum_{j=1}^{2}\Big(\langle n|E_{ij}|m\rangle-\langle n|F_{ij}|m\rangle-\langle n|G_{ij}|m\rangle+\langle n|H_{ij}|m\rangle\Big).
\end{equation}

Noted that we have assumed that the scalar field is in the thermal equilibrium state and its density matrix is
\begin{equation}
\label{eq:4.92a}
R_{0}=\prod_{\vec{k}}exp\Big(-\beta_{1}\omega_{k}A_{\vec{k}}^{\dag}A_{\vec{k}}\Big)\Big(1-exp(-\beta_{1}\omega_{k})\Big).
\end{equation}
Then it is easy to prove the following formulas:
\begin{equation}
\label{eq:4.93}
A_{\vec{k}}A_{-\vec{k}}A_{\vec{k}}^{\dag}A_{-\vec{k}}^{\dag}=A_{\vec{k}}^{\dag}A_{-\vec{k}}^{\dag}A_{\vec{k}}A_{-\vec{k}}+A_{\vec{k}}^{\dag}A_{\vec{k}}+A_{-\vec{k}}^{\dag}A_{-\vec{k}}+1;
\end{equation}
\begin{equation}
\label{eq:4.95}
\mathrm{Tr}_{B}(A_{\vec{k}}^{\dag}A_{\vec{k}}R_{0})=\frac{e^{-\beta_{1}\omega_{k}}}{1-e^{-\beta_{1}\omega_{k}}};
\end{equation}
\begin{equation}
\label{eq:4.96}
\mathrm{Tr}_{B}(A_{\vec{k}}^{\dag}A_{-\vec{k}}^{\dag}A_{\vec{k}}A_{-\vec{k}}R_{0})=\Big(\frac{e^{-\beta_{1}\omega_{k}}}{1-e^{-\beta_{1}\omega_{k}}}\Big)^{2}.
\end{equation}
\eqref{eq:4.95} is the average scalar particle number with the momentum $\vec{k}$. \eqref{eq:4.96} is the average particle pair number where this pair of particles has the momentum $\vec{k}$ and $-\vec{k}$, respectively.  In addition, we approximately perform the following two kinds of integration:
\begin{equation}
\label{eq:4.97}
\int_{0}^{\infty}ds e^{i(\alpha-\beta)s}=\pi\delta(\alpha-\beta)+i\mathbf{P}\frac{1}{\alpha-\beta}=\pi\delta(\alpha-\beta);
\end{equation}
\begin{equation}
\label{eq:4.98}
exp\Big\{i\int_{0}^{s}\omega_{k}dt\Big\}\approx exp\Big\{i\omega_{k}s+\varepsilon(s)\Big\}\approx e^{i\omega_{k}s}.
\end{equation}
Here, $\varepsilon(s)$ is a small quantity. Equation \eqref{eq:4.98} means that we approximately set $\omega_{k}$ as being unchanged with time. This is reasonable as we assumed that $\dot{a}$ is small. In \eqref{eq:4.97}, as usually done, we neglected the principle integral terms~\cite{SC}.

Using these equations from \eqref{eq:4.93} to \eqref{eq:4.98}, we can calculate all the matrix elements on the right hand side of equation \eqref{eq:4.92}. After we finish the calculation of these matrix elements, bringing all of them into \eqref{eq:4.92}, for the steady state, $\langle n|\dot{\rho}|m\rangle=0$, then we have
\begin{eqnarray}\begin{split}
\label{eq:4.99}
&\quad\big(\mathcal{C}\rho_{n-2,m}+\mathcal{B}\rho_{n,m+2}-\mathcal{A}\rho_{n-1,m+1}\big)e^{4i\langle\mathcal{H}\rangle t}\\
&+\big(\mathcal{C}\rho_{n,m-2}+\mathcal{B}\rho_{n+2,m}-\mathcal{A}\rho_{n+1,m-1}\big)e^{-4i\langle\mathcal{H}\rangle t}\\
&-2\mathcal{A}\rho_{n,m}+2\mathcal{C}\rho_{n-1,m-1}+2\mathcal{B}\rho_{n+1,m+1}=0.
\end{split}
\end{eqnarray}
Here
\begin{equation}
\label{eq:4.100}
\mathcal{B}=\Big(\frac{1}{e^{-\beta_{1}\langle\mathcal{H}\rangle}-1}\Big)^{2},
\end{equation}
\begin{equation}
\label{eq:4.101}
\mathcal{C}=1+\frac{2}{e^{-\beta_{1}\langle\mathcal{H}\rangle}-1}+\Big(\frac{1}{e^{-\beta_{1}\langle\mathcal{H}\rangle}-1}\Big)^{2},
\end{equation}
\begin{equation}
\label{eq:4.102}
\mathcal{A}=1+\frac{2}{e^{-\beta_{1}\langle\mathcal{H}\rangle}-1}+\frac{2}{(e^{-\beta_{1}\langle\mathcal{H}\rangle}-1)^{2}}.
\end{equation}

Obviously, the quantity $\mathcal{A}$, $\mathcal{B}$ and $\mathcal{C}$ have the following relations:
\begin{equation}
\label{eq:4.103}
\mathcal{A}>\mathcal{C}>\mathcal{B};
\end{equation}
\begin{equation}
\label{eq:4.104}
\mathcal{A}=\mathcal{B}+\mathcal{C}.
\end{equation}
Noted that the density matrix satisfy $\rho^{\dag}=\rho$. Therefore, we just need to solve the equation \eqref{eq:4.99} for the case of $m\geq n$. Setting $m=n+k$, where, $k=0,1,2,3,...$, and defining $f(n,k)=-\mathcal{A}\rho_{n,n+k}+\mathcal{C}\rho_{n-1,n+k-1}+\mathcal{B}\rho_{n+1,n+k+1}$, then equation \eqref{eq:4.99} becomes
\begin{equation}
\label{eq:4.104a}
f(n-1,k+2)e^{4i\langle\mathcal{H}\rangle t}+f(n+1,k-2)e^{-4i\langle\mathcal{H}\rangle t}+2f(n,k)=0.
\end{equation}
Equation \eqref{eq:4.104a} is valid for any value of $n$ ($n=1,2,3,...$) and $k$ ($k=0,1,2,3...$). One can easily verify that $f(n,k)=0$ ($f(n,k)=0$ means that $f(n-1,k+2)=0$ and $f(n+1,k-2)=0$ ) is the solution of equation \eqref{eq:4.104a}. Then equation \eqref{eq:4.99} can be reduced to
\begin{equation}
\label{eq:4.105}
-\mathcal{A}\rho_{n,n+k}+\mathcal{C}\rho_{n-1,n+k-1}+\mathcal{B}\rho_{n+1,n+k+1}=0.
\end{equation}

Solving equation \eqref{eq:4.105}, we can obtain the steady state solution ( more strictly, the quasi-steady state solution, see the later discussions for details ) of the quantum master equation \eqref{eq:4.89}:
\begin{equation}
\label{eq:4.106}
\rho_{n,n+k}=(\frac{\mathcal{C}}{\mathcal{B}})^{n-2}\cdot \frac{\mathcal{A}}{\mathcal{B}}\cdot\rho_{1,k+1}+\frac{\mathcal{B}}{\mathcal{C}-\mathcal{B}}\cdot\rho_{1,k+1}\cdot\Big\{(\frac{\mathcal{C}}{\mathcal{B}})^{n-2}-1\Big\}.
\end{equation}
Particularly, when k=0, setting $P(n)=\rho_{n,n}$, we have
\begin{equation}
\label{eq:4.107}
P(n)=(\frac{\mathcal{C}}{\mathcal{B}})^{n-2}\cdot \frac{\mathcal{A}}{\mathcal{B}}\cdot P(1)+\frac{\mathcal{B}}{\mathcal{C}-\mathcal{B}}\cdot P(1)\cdot\Big\{(\frac{\mathcal{C}}{\mathcal{B}})^{n-2}-1\Big\}.
\end{equation}
\eqref{eq:4.107} is the distribution of the diagonal element of the reduced density matrix $\rho$. $P(n)$ represents the fraction of the state $|n\rangle$. If we measure the volume of the space, then we have the probability $P(n)$ to observe the volume of the space being $2n$. $P(1)$ can be fixed by the normalization condition.

We point out that the uniqueness of the steady state solution \eqref{eq:4.106} is not easy to strictly prove. Roughly speaking, if the gravitational system is weakly coupled to the thermal bath, then it will reach the equilibrium state after the transient relaxation. And the equilibrium state is the steady state. This is the common characteristic of the open system which weakly coupled to only one thermal bath. The distribution \eqref{eq:4.107} is the equilibrium distribution (see the next subsection for details). This may indicate that the solution \eqref{eq:4.106} is unique.

According to the classical Friedmann equation, as long as the energy density of the matter is not equal to zero, the universe will expand or contract.  In our model, the energy density of the bath is obviously greater than zero, thus one may feel strange about the solution of equation \eqref{eq:4.106} which seems to be a steady state. We should point out that although equation \eqref{eq:4.106} strictly satisfy $d\rho_{n+k}/dt=0$, we can not think of \eqref{eq:4.106} as representing a strictly steady state of the system. This is because that in equation \eqref{eq:4.91}, we constrain that the temperature of the bath is approximately not changed in a enough long period of time (longer than the relaxation time scale of the system). This is reasonable as we constrain that the variation rate of the classical scale factor $\dot{a}$ is small. Then we can also approximately treat the parameters $\mathcal{A}$, $\mathcal{B}$, $\mathcal{C}$ and $\mathcal{H}$ as not being changed significantly in a enough long period of time. Only under this kind of approximation, the quantum master equation \eqref{eq:4.91} has the solution \eqref{eq:4.106}. However, if we consider a very long time interval so that the variation of the bath temperature can not be neglected, in this case, we can not use equation \eqref{eq:4.91} to describe the evolution of the system. Thus, more precisely, we say the solution \eqref{eq:4.106} is a quasi-steady state. The meaning of the quasi-steady state is: after a short time relaxation, the system will reach this state. In the relaxation process, the temperature of the bath can be approximately thought of not changing with time. This treatment is similar to the case of the radiation of the black hole. As the black hole can radiate particles to outside of the event horizon, the mass of the black hole decreases gradually. At the time scale of the event horizon radiate a particle, the variation of the black hole mass is very small, so that we can approximately treat the mass of the black hole as nearly constant when we derive the temperature of the black hole.

Although in \eqref{eq:4.89}, the upper limit of the integral is infinite, this does not mean the integral is carried out on the whole time interval of the evolution of the universe. This just means that the integral is carried out on an enough long period which the system could reach the quasi-steady state. As time goes by, the volume of the space will expand and become larger, the quantum features of the system will gradually become less important. Eventually, the system will become a classical one so that we can describe it by the Friedmann equations. Finally, the physical volume of the system will expand to be infinitely large. The temperature of the bath will approach to zero. However, the equation \eqref{eq:4.91} can not describe this interesting process. The reason is that in this process, the variation of the bath temperature or the effect of the back reaction from gravity to the scalar field can not be neglected. To find an equation which can describe this process is beyond the scope of this work.

\begin{figure}[tbp]
\centering
\subfigure[]{
\begin{minipage}[t]{0.48\textwidth}
\centering
\includegraphics[width=7.5cm]{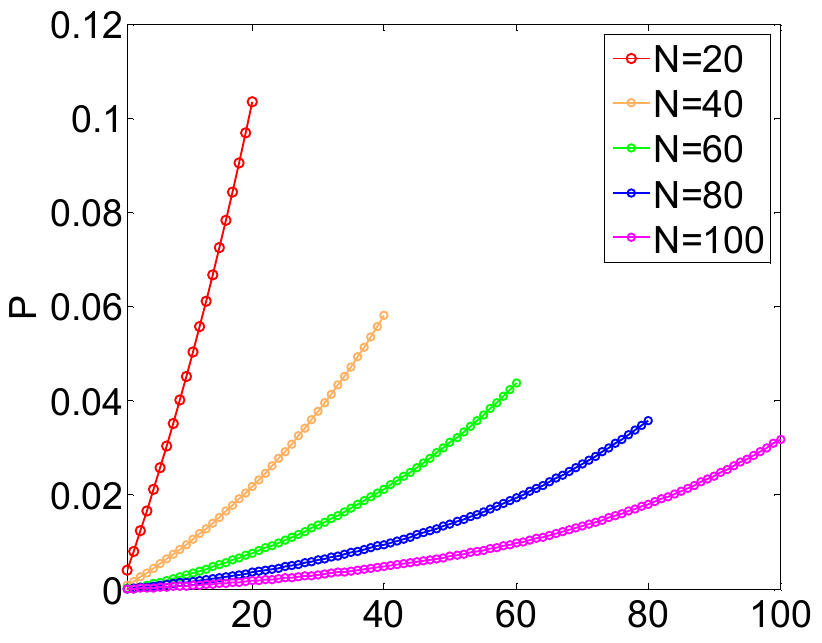}
\label{fig:a}
\end{minipage}}
\subfigure[]{
\begin{minipage}[t]{0.48\textwidth}
\centering
\includegraphics[width=7.5cm]{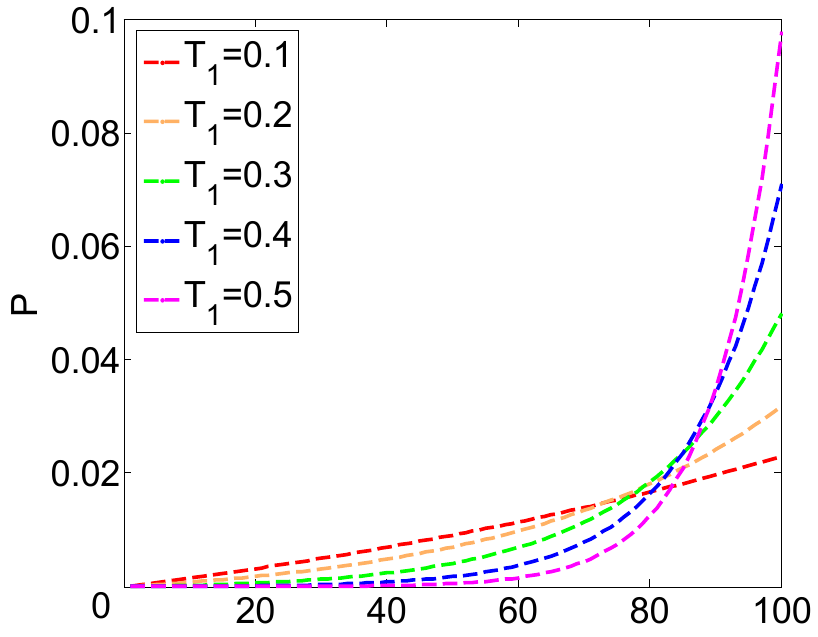}
\label{fig:b}
\end{minipage}}
\caption{Distribution of the diagonal elements of density matrix. The horizontal axis represents the serial number of the quantum state while the vertical axis represents the fraction of the related quantum state. In figure 3(a), the temperature of the bath is $T_{1}=0.2$. Different curves correspond to different total number of the quantum state. In figure 3(b), $N=100$, different curves are related to different temperatures.}
\label{fig:3}
\end{figure}

Setting $m_{1}=0.01$ and $\ell_{0}=(\gamma\sqrt{\triangle})^{1/3}$, then one can calculate the numerical values of the probability distribution $P(n)$, as shown in figure~\ref{fig:3}.  Suppose $N$ is the dimension of the Hilbert space. As the number of the eigenvector of the volume operator is infinite, $N$ is infinite. But as a toy model, we set $N$ as a finite number. In figure 3(a), different curves correspond to different $N$. From this figure, we learn that the shape of different curves are  similar. In all of these curves, $P(n)$ increases with the horizontal axis. Hence we have more opportunities to observe the space in the bigger volume quantum state. We expect that as $N$ becomes bigger, the curve still has the similar trend. For simplicity, we set $N=100$ both in figure 3(b) and in the later discussions. In figure 3(b), different curves are related to different bath temperatures. Figure 3(b) shows that when the bath temperature increases, the spacetime has higher probability or chance to stay in the larger volume state. The physical explanation is that when the environmental temperature is higher, the Hamiltonian of the bath is also larger. However, the total Hamiltonian of the space time and the environment is zero. Therefore, the Hamiltonian of the space time becomes more negative. This favors larger size universe. We will show in the next that the distribution presented in figure~\ref{fig:3} is an equilibrium state and the detailed balance is preserved.

\subsubsection{Quantum geometrical current and coherence}

For the ordinary open system, the current is an important physical quantity. In the non-steady state, the current can drive the system evolution. And in the non-equilibrium steady state, it is an important physical entity to quantify the irreversibility and the dissipation structure. If there is a time arrow in the system, then there exists the irreversible current~\cite{JW,JW2}. In this subsection, we will reveal that for the open quantum gravitation system, the current also has the similar function. This may provide us a different perspective about the evolution of the spacetime and can help to understand the time arrow of the universe.

For the non-steady state, taking $m=n$ in \eqref{eq:4.92}, after calculating all the elements on the right hand side of \eqref{eq:4.92}, we obtain
\begin{eqnarray}\begin{split}
\label{eq:4.114}
\frac{dP(n)}{dt}=&\pi\eta^{2}\langle\mathcal{H}\rangle^{2}\Big(\langle\mathcal{H}\rangle^{2}-m_{1}^{2}\Big)\Big\{-\mathcal{A}P(n)+\mathcal{C}P(n-1)+\mathcal{B}P(n+1)\\
&+\Big(2\mathcal{C}\rho_{n-2,n}+2\mathcal{B}\rho_{n,n+2}-2\mathcal{A}\rho_{n-1,n+1}\Big)\cdot e^{4i\langle\mathcal{H}\rangle t}+h.c.n.\Big\}\neq0,
\end{split}
\end{eqnarray}
where $h.c.n.$ represents the Hermitian conjugate of all the non-diagonal terms in \eqref{eq:4.114} and
\begin{equation}
\label{eq:4.115}
\eta=\frac{6\pi V(0)\xi}{\gamma \ell_{0}^{3}\sqrt{\Delta}}.
\end{equation}
From equation \eqref{eq:4.99} and \eqref{eq:4.114}, one can see that the global factor $\pi\eta^{2}\langle\mathcal{H}\rangle^{2}(\langle\mathcal{H}\rangle^{2}-m_{1}^{2})$ has no  influence on the quasi-steady state. This factor can influence the variation rate of the reduced density matrix. The larger global factor corresponds to the faster variation of the system.

From equation \eqref{eq:4.114}, we can see that the transition rate from the state $|n-1\rangle$ to the state $|n\rangle$ is $\pi\eta^{2}\langle\mathcal{H}\rangle^{2}(\langle\mathcal{H}\rangle^{2}-m_{1}^{2})\mathcal{C}$. And the transition rate from the state $|n+1\rangle$ to the state $|n\rangle$ is $\pi\eta^{2}\langle\mathcal{H}\rangle^{2}(\langle\mathcal{H}\rangle^{2}-m_{1}^{2})\mathcal{B}$. Thus if we use $\Gamma_{ij}$  to represent the transition rate from the state $|i\rangle$ to the state $|j\rangle$, then we have
\begin{equation}
\label{eq:4.109}
  \Gamma_{ij}=\begin{cases}
      \pi\eta^{2}\langle\mathcal{H}\rangle^{2}(\langle\mathcal{H}\rangle^{2}-m_{1}^{2})\mathcal{B}\,;\quad j=i-1\\
     \pi\eta^{2}\langle\mathcal{H}\rangle^{2}(\langle\mathcal{H}\rangle^{2}-m_{1}^{2})\mathcal{C}\,;\quad j=i+1\\
     0\,; \quad others
   \end{cases}.
\end{equation}

Define~\cite{JW,HW}:
\begin{equation}
\label{eq:4.110}
F_{mn}=P(m)\Gamma_{mn}-P(n)\Gamma_{nm}.
\end{equation}
For the quasi-steady state ($dP(n)/dt=0$), combining \eqref{eq:4.109} and \eqref{eq:4.110}, after carrying out the normalization for $P(n)$, then we obtain
\begin{equation}
\label{eq:4.111}
F_{mn}\approx 0.
\end{equation}
In \eqref{eq:4.111}, $F_{mn}$ is just approximately equal to zero as we set the dimension of the Hilbert space $N$ to be a finite number. If we choose $N$ as infinity, $F_{mn}$ will be strictly equal to zero. The physical meaning of $F_{mn}$ is the variation rate of the state $|n\rangle$ induced by the transition between the state $|m\rangle$ and $|n\rangle$. Intuitively, $F_{mn}$ represents the current from the state $|m\rangle$ to the state $|n\rangle$. $F_{mn}=0$ means that the detailed balance is being preserved. The gravitational system is in equilibrium with the scalar field bath.

One often uses the entropy production rate (EPR) to measure the time irreversibility of the system. It is defined by~\cite{JS}
\begin{equation}
\label{eq:4.111a}
EPR=\sum_{ij}P(i)\Gamma_{ij}\textrm{ln}\frac{P(i)\Gamma_{ij}}{P(j)\Gamma_{ji}}.
\end{equation}
From \eqref{eq:4.110} and \eqref{eq:4.111a}, we can see that the relationship between the current and the EPR is
\begin{equation}
\label{eq:4.111b}
EPR=\frac{1}{2}\sum_{ij}F_{ij}\textrm{ln}\frac{P(i)\Gamma_{ij}}{P(j)\Gamma_{ji}}.
\end{equation}
Combining \eqref{eq:4.111} and \eqref{eq:4.111b}, we learn that in the quasi-steady state, the EPR of the gravitational system is zero. Thus, there is no irreversibility.

For the non-steady state, in general $F_{mn}\neq0$. The EPR is also in general not equal to zero. According to \eqref{eq:4.110}, obviously, we have $F_{mn}=-F_{nm}$. Since both $F_{mn}$ and $F_{nm}$ represent the same current between the state $|m\rangle$ and $|n\rangle$. For convenience, we ignore the one which is smaller than zero to reach the following definition~\cite{MQ,HW}
\begin{equation}
\label{eq:4.112}
J_{ij}=P(i)\Gamma_{ij}-min\{P(i)\Gamma_{ij},P(j)\Gamma_{ji}\}.
\end{equation}
Because $\{|i\rangle\}$ represents a set of quantum geometry states, we can term $J_{ij}$ (or $F_{ij}$) as the quantum geometry current.  If the space transition is from the state $|i\rangle$ to the state $|j\rangle$, the volume of the space would change $\Delta V=V_{j}-V_{i}$, where $V_{i}$ ($V_{j}$) is the space volume in the state $|i\rangle$ ($|j\rangle$). Thus the current can induce the variation of the space volume, the variation rate induced by the current $F_{ij}$ is
\begin{equation}
\label{eq:4.113}
R_{i,j}=V_{j}F_{ij}.
\end{equation}

Combining \eqref{eq:4.109}, \eqref{eq:4.110} and \eqref{eq:4.114},  we have
\begin{eqnarray}\begin{split}
\label{eq:4.116}
\frac{dP(n)}{dt}=&\sum_{m}F_{mn}+\pi\eta^{2}\langle\mathcal{H}\rangle^{2}\Big(\langle\mathcal{H}\rangle^{2}-m_{1}^{2}\Big)\Big\{\Big(2\mathcal{C}\rho_{n-2,n}+2\mathcal{B}\rho_{n,n+2}\\&-2\mathcal{A}\rho_{n-1,n+1}\Big)\cdot e^{4i\langle\mathcal{H}\rangle t}+h.c.n.\Big\}.
\end{split}
\end{eqnarray}
On the other hand, the variation rate of the average volume is given as
\begin{equation}
\label{eq:4.117}
\frac{d\langle V\rangle}{dt}=\mathrm{Tr}(V\frac{d\rho}{dt})=\sum_{n}V_{n}\frac{dP(n)}{dt}.
\end{equation}
Substituting \eqref{eq:4.116} into \eqref{eq:4.117}, we reach
\begin{eqnarray}\begin{split}
\label{eq:4.118}
\frac{d\langle V\rangle}{dt}=&\sum_{mn}V_{n}F_{mn}+\pi\eta^{2}\langle\mathcal{H}\rangle^{2}\Big(\langle\mathcal{H}\rangle^{2}-m_{1}^{2}\Big)\sum_{n}V_{n}\Big\{\Big(2\mathcal{C}\rho_{n-2,n}+2\mathcal{B}\rho_{n,n+2}\\&-2\mathcal{A}\rho_{n-1,n+1}\Big)\cdot e^{4i\langle\mathcal{H}\rangle t}+h.c.n.\Big\}.
\end{split}
\end{eqnarray}

Equation \eqref{eq:4.118} clearly shows that the current $F_{mn}$ can drive the evolution of the average value of the space volume. Although we derived this result in a toy model, it is easy to see the logic from \eqref{eq:4.114} to \eqref{eq:4.118} is very general. Whether the bath temperature is changed with time or not, equation \eqref{eq:4.117} is generally correct.

Not only the variation rate of the average volume can be driven by this current, for the general geometry operator which is formally represented by $\hat{G}$, the variation rate of the average value of $\hat{G}$ can also be driven by this current. In \eqref{eq:4.117} and \eqref{eq:4.118}, we can replace $V$ by $\hat{G}$, then we can immediately see that the variation rate of the average value of the operator $\hat{G}$ is driven by this current. This in fact provides us a new viewpoint about the evolution of the geometry.

In addition to the current $F_{mn}$, there are other non-diagonal terms in \eqref{eq:4.118}. The non-diagonal term of the density matrix $\rho$ represents the coherence. Therefore, equation \eqref{eq:4.118} shows that the coherence also drives the evolution of the geometry. When the non-diagonal terms of the density matrix are approximately equal to zero so that the coherence is not important, the evolution of the spacetime is only driven by the current $F_{mn}$. In this case, equation \eqref{eq:4.116} and \eqref{eq:4.118} become
\begin{equation}
\label{eq:4.118b}
\frac{dP(n)}{dt}=\sum_{m}F_{mn}
\end{equation}
and
\begin{equation}
\label{eq:4.118c}
\frac{d\langle V\rangle}{dt}=\sum_{mn}V_{n}F_{mn},
\end{equation}
respectively. Equation \eqref{eq:4.118b} is the Pauli master equation. Equation \eqref{eq:4.118c} is the evolution equation of $\langle \hat{V}\rangle$ while the evolution of $P(n)$ is governed by the Pauli master equation \eqref{eq:4.118b}. It is clear that equations \eqref{eq:4.116} and \eqref{eq:4.118} are different from equations \eqref{eq:4.118b} and \eqref{eq:4.118c}. In equation \eqref{eq:4.118b} and \eqref{eq:4.118c}, there are no non-diagonal terms. However, for the quantum system, usually the coherence is important and the non-diagonal terms in equation \eqref{eq:4.116} and \eqref{eq:4.118} can not be neglected. Thus for the quantum gravitational system, both the current $F_{mn}$ and the coherence drive the evolution of the quantum geometry.

In general, the dynamics of the open quantum system is described by the generalized Lindblad equation~\cite{IVD,HEJB}. The classical limit of the generalized Lindblad equation is the Pauli master equation which can be written as $dP(i)/dt=\sum_{j}F_{ji}$~\cite{HEJB,HW}. Thus, in the classical limit, the dynamics of the system is determined by the current. For the quantum system, the coherence terms in the generalized Lindblad equation are important. The dynamics of the quantum system is determined  by both the current and the coherence~\cite{JW3,JW4}. For the situation of the quantum gravity, after tracing out the matter fields, the dynamics of the quantum spacetime should also be described by the generalized Lindblad equation. Thus it is reasonable to infer that the dynamics of the quantum spacetime is driven by the current and the coherence in more general situations, not just in our toy model.

\begin{figure}[tbp]
\centering
\includegraphics[width=7.5cm]{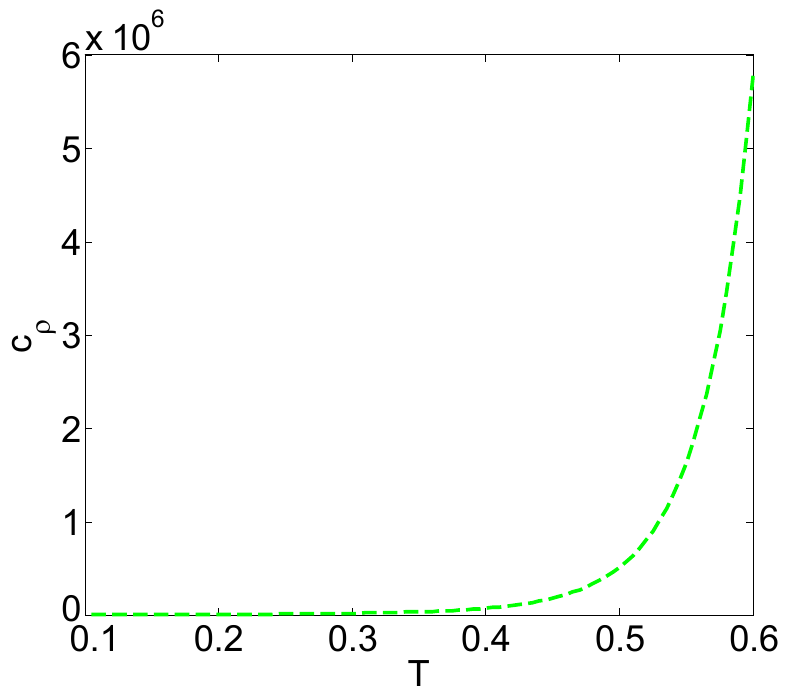}
\caption{\label{fig:4} The coherence versus the variation of the temperature of the bath. The horizontal axis represents the temperature of the bath. The vertical axis represents the coherence ($c_{\rho}$).}
\end{figure}

As a quantum system, there is coherence in the quasi-steady state. We can use the following definition to measure the coherence of the reduced gravitation system~\cite{TM}:
\begin{equation}
\label{eq:4.125}
c_{\rho}=\sum_{mn}\rho_{mn}-\sum_{n}\rho_{nn}.
\end{equation}
Figure~\ref{fig:4} shows the variation of the coherence with the temperature of the bath. In figure~\ref{fig:4}, we still fix the parameters $m_{1}=0.01$  and $\ell_{0}=(\gamma\sqrt{\triangle})^{1/3}$.  In order to calculate the coherence based on the definition \eqref{eq:4.125}, we need to fix the value of the density matrix element $\rho_{1,k+1}$ in \eqref{eq:4.106}. We choose $\rho_{1,k+1}=0.01$. Figure~\ref{fig:4} shows when the temperature of the bath increases, the coherence of the gravitation subsystem also increases. This result is consistent with the intuition: at the very early stage, the universe is in quantum state being hot with strong quantum coherence. Later on, the universe becomes cold and classical, thus looses the coherence.

\subsection{The continuous limit}
When the smallest element of the area $\Delta=4\pi\gamma\sqrt{3}$ goes to zero, the loop quantum cosmology can approach to the usual canonical quantum cosmology which is described by the Wheeler-DeWitt equation~\cite{AAP}. The eigenvalue spectrum of the volume operator tends to be a continuous spectrum. That is,
\begin{equation}
\label{eq:4.126}
V_{n}=2\pi\gamma\sqrt{\Delta}\cdot \mathcal{V}_{n}=4\pi\gamma\sqrt{\Delta}\cdot n\longrightarrow V_{a}=a^{3}\ell_{0}^{3},
\end{equation}
where $a$ is the scale factor in the FRW metric and $V_{a}$ represents the physical volume of the space. Therefore, the diagonal elements of the density matrix in equation \eqref{eq:4.107} should become:
\begin{equation}
\label{eq:4.127}
P(a)=\lim_{\Delta \to 0}\Big\{(\frac{\mathcal{C}}{\mathcal{B}})^{\frac{a^{3}}{4\pi\gamma\sqrt{\Delta}}}\cdot \frac{\mathcal{A}}{\mathcal{B}}\cdot P(a=0)+\frac{\mathcal{B}}{\mathcal{C}-\mathcal{B}}\cdot P(a=0)\cdot\Big\{(\frac{\mathcal{C}}{\mathcal{B}})^{\frac{a^{3}}{4\pi\gamma\sqrt{\Delta}}}-1\Big\}\Big\}.
\end{equation}
But in equation \eqref{eq:4.127}, $\lim_{a\to0}P(a)\neq P(a=0)$. This seems unreasonable as it is natural to think that $P(a)$  as a continuous function.

Noted that $P(a=0)$ and $P(a=0)\mathcal{A}/\mathcal{B}$ are related to the state $|1\rangle$ and $|2\rangle$ in the discrete case, respectively. After taking the continuous limit, in order to ensure that $P(a)$ is a continuous function, we must require that the probability difference of these two states is infinitely close to zero. For this purpose, we only need to replace $P(a=0)\mathcal{A}/\mathcal{B}$ by $P(a)$ in \eqref{eq:4.127}. Thus we can modify $P(a)$ in \eqref{eq:4.127} as
\begin{equation}
\label{eq:4.128}
P_{Q}(a)=\lim_{\Delta \to 0}\Big\{(\frac{\mathcal{C}}{\mathcal{B}})^{\frac{a^{3}}{4\pi\gamma\sqrt{\Delta}}}\cdot P(a=0)+\frac{\mathcal{B}}{\mathcal{C}-\mathcal{B}}\cdot P(a=0)\cdot\Big\{(\frac{\mathcal{C}}{\mathcal{B}})^{\frac{a^{3}}{4\pi\gamma\sqrt{\Delta}}}-1\Big\}\Big\}.
\end{equation}
In loop quantum cosmology, the zero volume state can not evolve into the finite volume state and vice versa~\cite{AE,ATP,ATPP,AATP}. But in the canonical quantum cosmology, this can happen. Similarly, in the continuous limit, the non-diagonal elements of the density matrix in equation \eqref{eq:4.106} become
\begin{equation}
\label{eq:4.129}
\rho_{a,a'}=\lim_{\Delta \to 0}\Big\{(\frac{\mathcal{C}}{\mathcal{B}})^{\frac{a^{3}}{4\pi\gamma\sqrt{\Delta}}}\cdot \rho_{a=0,a'}+\frac{\mathcal{B}}{\mathcal{C}-\mathcal{B}}\cdot\rho_{a=0,a'}\cdot\Big\{(\frac{\mathcal{C}}{\mathcal{B}})^{\frac{a^{3}}{4\pi\gamma\sqrt{\Delta}}}-1\Big\}\Big\}.
\end{equation}

The distribution  $P_{Q}(a)$ in \eqref{eq:4.128} does not explicitly depend on the initial state of the spacetime. This is the feature of the quantum master equation.  $P_{Q}(a)$ also represents the transition probability from certain initial state to the state $|a\rangle$ (the space volume is definitely equal to $a^{3}l_{0}^{3}$). However, as we constrain the temperature of the bath as being approximately not significantly changed with time, the difference of the average value of the space volume between the initial state and the final state should be small. The difference between the initial state and the final state is that the initial state has larger fluctuations, yet the final state is in equilibrium with the bath and thus has smaller fluctuations. Thus $P_{Q}(a)$ does not represent the probability of the universe tunneling from the zero volume state to a finite volume state.  $P_{Q}(a)$ can then be used to represent the transition (or approximately the survival) probability in the presence of the fluctuations from the initial state to the state $|a\rangle$ (the scale factor is $a$).  This is different with the Hartle-Hawking and the Vilenkin wave function.

When $\dot{a}$ is small or $\dot{a}^{2}=a^{2}H^{2}\ll 1$, The Hartle-Hawking wave function of the universe gives $\psi_{H-H}=e^{a^{2}/2}$~\cite{JH}. The Vilenkin wave function of the universe is $\psi_{V}=e^{-a^{2}/2}$~\cite{AV,AVV}. Both $\psi_{H-H}$ and $\psi_{V}$ represent the tunneling amplitude of the universe from the zero volume state to a finite volume state. They are different as the boundary condition is different. The tunneling probability distribution corresponding to $\psi_{H-H}$ and $\psi_{V}$ are
\begin{equation}
\label{eq:4.130}
P_{H-H}=\frac{e^{a^{2}}}{Z_{1}}
\end{equation}
and
\begin{equation}
\label{eq:4.131}
P_{V}=\frac{e^{-a^{2}}}{Z_{2}},
\end{equation}
respectively. Here, $Z_{1}$ and $Z_{2}$ are the partition function.

\begin{figure}[tbp]
\centering
\subfigure[]{
\begin{minipage}[t]{0.48\textwidth}
\centering
\includegraphics[width=7.5cm]{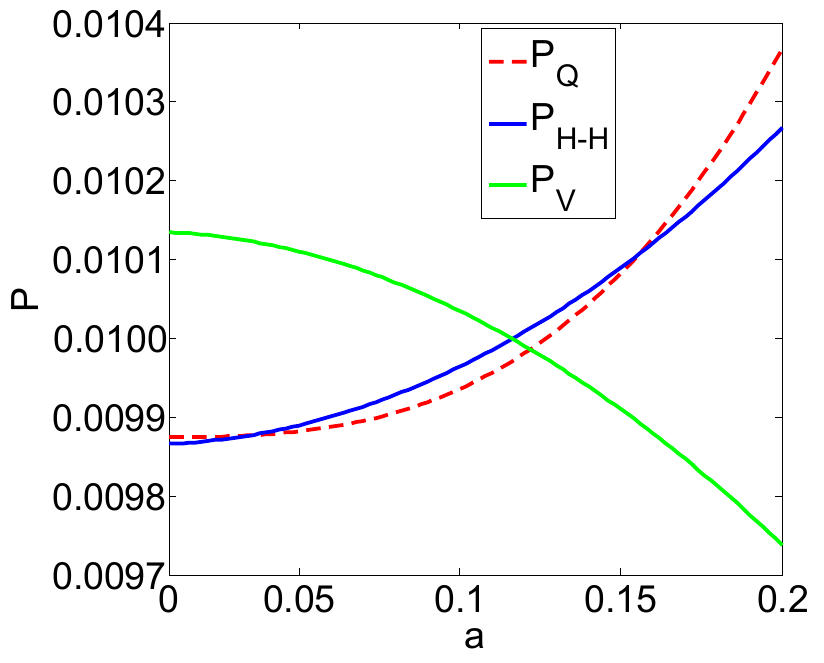}
\label{fig:5(a)}
\end{minipage}}
\subfigure[]{
\begin{minipage}[t]{0.48\textwidth}
\centering
\includegraphics[width=7.5cm]{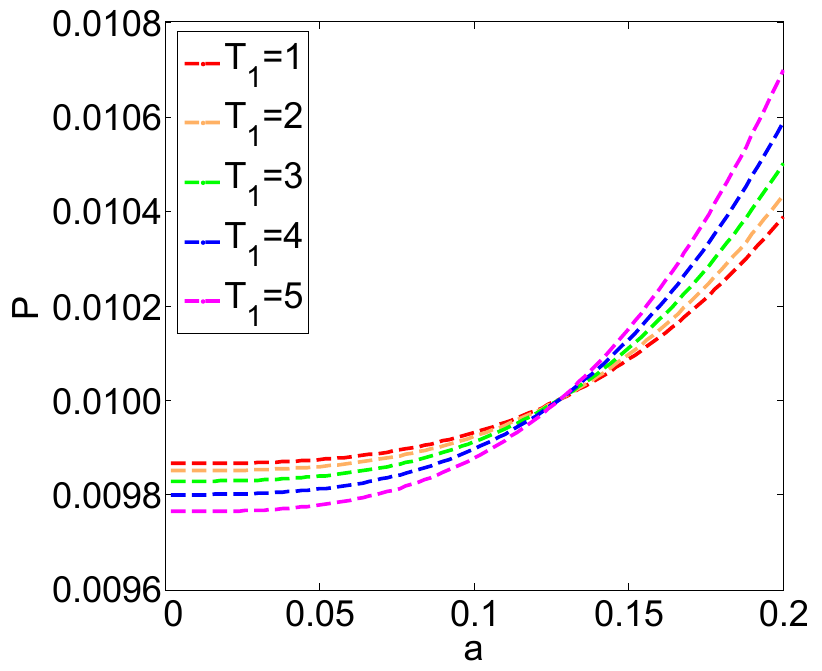}
\label{fig:5(b)}
\end{minipage}}
\caption{The probability distribution. The horizontal axis represents the scale factor in the FRW metric and the vertical axis represents the probability.}
\label{fig:5}
\end{figure}
$P_{Q}$, $P_{H-H}$ and $P_{V}$ are shown in figure~\ref{fig:5(a)}. In this figure, we set $m_{1}=0.01$, $T_{1}=0.2$ and $\ell_{0}=1$. The red dotted curve represents $P_{Q}$, the blue solid curve represents $P_{H-H}$ and the green solid curve represents $P_{V}$. From figure~\ref{fig:5(a)}, we learn that both $P_{Q}$ and $P_{H-H}$ monotonically increase with the variable $a$, $P_{V}$  monotonically decreases. They are different for certain reasons. First, both $P_{H-H}$ and $P_{V}$ are related to the action
\begin{equation}
\label{eq:4.132}
S=\frac{1}{16\pi}\int dx^{4}\sqrt{-g}R+\frac{1}{2}\int dx^{4} \sqrt{-g}g^{\mu\nu}\phi_{2,\mu}\phi_{2,\nu}
-\frac{1}{2}\int dx^{4} \sqrt{-g}(m_{2}^{2}\phi_{2}^{2}+\frac{1}{2}\lambda\phi_{2}^{4}).
\end{equation}
But $P_{Q}$ is related to the action \eqref{eq:4.1}. In \eqref{eq:4.132}, the scalar field is minimally coupled with the gravity. But in \eqref{eq:4.1}, the interaction between the scalar field and the gravity is not minimally coupled. Second, the boundary condition is different. $P_{H-H}$ corresponds to the so called "no-boundary" boundary condition where there is no spacetime and matter initially. $P_{V}$ corresponds to the so called tunneling boundary condition where there exist matter and the space volume is zero. $P_{Q}$ corresponds to the boundary condition where the scalar field is in the thermal state and the information of the initial state of the spacetime is not relevant for $P_{Q}$. Third, the approximation method is different. $P_{H-H}$ and $P_{V}$ are related to the semi-classical approximation, but $P_{Q}$ is related to the Born-Markov approximation. Fourth, $P_{H-H}$ and $P_{V}$ correspond to the FRW metric where the space slice is curved ($k=1$), but $P_{Q}$ as the transition probability corresponds to the FRW metric where the space slice is flat ($k=0$).

In addition, noted that in the region where the scale factor $a$ and the mode amplitudes of the scalar field are small, for the Vilenkin wave function, the tunneling amplitude of the universe from the zero volume state to the finite volume state is nearly not influenced by the non-minimal coupling term~\cite{SWM}. Thus , $P_{V}$ not only represents the tunneling amplitude of the universe in the case of the minimal coupling, but it also can be used to describe the tunneling amplitude of the universe from  zero volume state to small volume state in the case of the non-minimal coupling.

In figure~\ref{fig:5(b)}, we set $m_{1}=0.1$ and $l_{0}=1$. In this figure, different curves correspond to different bath temperatures. From this figure, we can see that when the bath temperature increases, the spacetime has higher probability being in the larger volume state. We find that figure~\ref{fig:5(b)} and figure~\ref{fig:b} are similar to each other. Thus, while the parameter $\Delta$ tends toward to zero, the influence of the bath temperature to the probability distribution is qualitatively not changed.

\begin{figure}[tbp]
\centering
\includegraphics[width=9cm]{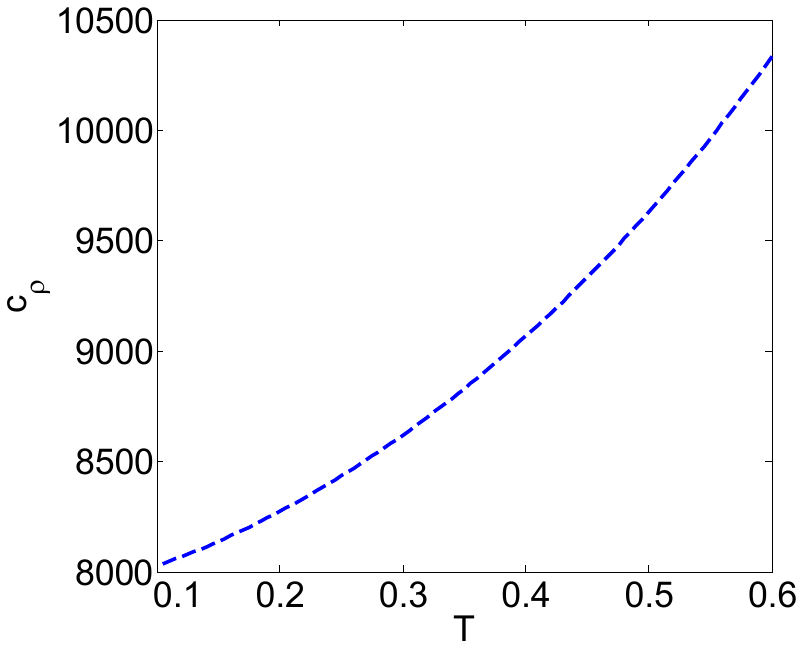}
\caption{\label{fig:6} Coherence in the continuous limit. The horizontal axis represents the temperature of the bath and the vertical axis represents the coherence.}
\end{figure}

Figure~\ref{fig:6} shows the variation of the coherence of the reduced gravitation system with the temperature of the bath in the continuous limit. We set $m_{1}=0.01$, $\ell_{0}=1$, $\rho_{a=0,a'}=0.01$ and constrain the scale factor $a\leq 1$. In this figure, the horizontal axis and the vertical axis represent the temperature of the bath and the coherence of the spacetime, respectively. We can see from figure~\ref{fig:6} that the coherence  monotonically decreases as the temperature of the bath decreases. This is similar to the discreteness case.

If the spacetime is dominated by the non-relativistic matter, then the higher temperature of the bath is usually related to the smaller volume of the space. This corresponds to the earlier times of our universe. Figure~\ref{fig:6} indicates that at the early times of the universe, the coherence is important. That is to say, as our universe expand to become bigger, the temperature of the bath decreases, thus the coherence decreases. Therefore, the expanding of our universe leads to the less coherence. This imply a quantum universe can expand to a classical one. This is also similar to the black hole evaporation process. As the black hole radiates particles, the mass of the black hole decrease. This can lead the quantum effect of the black hole to become more and more important. At the same time, due to the  heat capacity of the black hole being negative, the temperature of the black hole will increase. Thus, for the black hole, it appears that the coherence also monotonically increases with the temperature.

\section{Conclusions}
In this work, at first we developed a parameterized theory for the open quantum gravitation system, that is equation \eqref{eq:2.34}. Based on this equation, one can study the evolution of the density matrix along the bubble time. The evolution along the bubble time is determined by the Super-Hamiltonian vector. Equation \eqref{eq:2.34} can be used for the non-inertial frame. One of a smeared version of  equation \eqref{eq:2.34} is equation \eqref{eq:2.37}. Equation \eqref{eq:2.37} can also be strictly derived by introducing the Brown-Kucha$\check{\mathrm{r}}$  dust field~\cite{JD,VT,HWJ}.

The general covariance requires the Hamiltonian to be zero for any isolated system~\cite{CK}. This leads to the difficulty to understand the evolution of the universe. This problem is still unsolved up to now~\cite{AK,SV,MY}. We can divide the total isolated system into two parts: a subsystem plus the environment. Usually, as the subsystem has interaction with the environment, the subsystem should be seen as an open system. The general covariance does not require the Hamiltonian of the subsystem to be zero, so the general covariance does not require the time derivative of the density matrix of the subsystem to be zero. On the other hand, due to the subsystem coupling with the environment, usually the entanglement entropy and the energy of the subsystem will change with time. This clearly indicates that the diagonal element of the density matrix of the subsystem will change with time. The time derivative of the density matrix in \eqref{eq:2.37} then should not be equal to zero. We expect that the equation \eqref{eq:2.37} can be used to describe the evolution of the density matrix along the coordinate time.  This may provide insights for solving the time problem in quantum gravity.

Based on equation \eqref{eq:2.37} and using the Born-Markov approximation~\cite{HF,HC}, one can obtain the quantum master equation \eqref{eq:3.7}. Although equation \eqref{eq:2.37} can be used to describe any open quantum gravitation system, equation \eqref{eq:3.7} can only be used to describe certain specific gravitation systems. Equation \eqref{eq:3.7} can be used to describe certain interesting properties of the open quantum gravitation system, such as the non-unitary evolution, the decoherence and the nonequilibrium evolution.

Finally, we studied a specific model where the real scalar field plays the role of the bath. We considered the simple case where the evolution of the geometry is very slow and the spacetime is homogenous and isotropic. Usually, the quantization of the field in curved spacetime is difficult~\cite{RW}. If the cosmological particle production can be neglected, we can still approximately introduce the creation and annihilation operators for the field over the entire period of time. As a toy model, although it can not be used to describe exactly how our universe evolves, it still can reveal some interesting features.

We obtained the quasi-steady state solution of the quantum master equation for this model. The solution indicates that the space has larger probability in the bigger volume state. We found that in the quasi-steady state, the spacetime geometry is in the equilibrium state, the detailed balance is preserved and the entropy production rate is zero. In the non-steady state, in general, the scalar field can induce the emergence of the non-zero quantum geometry current. This current is defined in the volume representation, so we call it quantum geometry current. This is different with the conventional heat flux which is defined in the particle number representation. We showed that this current and the coherence can drive the evolution of the spacetime geometry.  We also show that the coherence of the reduced gravitation system increases when the temperature of the bath increases.

For the non-steady state, the variation rate of the average volume is not equal to zero. \eqref{eq:4.118} clearly shows that the variation of the average volume is driven by the current and the coherence. The coherence can also drive the evolution of the geometry.  This is consistent with the fact in the quantum thermodynamics that one can extract work from the coherence. This provides us a new view about the evolution of the spacetime geometry.

After taking the continuous limit, we show that the diagonal elements of the density matrix can represent the transition probability of the universe from certain initial  state to the final state (the eigenstate of the volume operator). There are also other ways to study the tunneling of the universe in the framework of the loop quantum cosmology~\cite{SDM,MSA}. We show that when the bath temperature increases, the universe has higher probability being in the larger volume state.

\section*{Acknowledgements}
Hong Wang thanks for support from National Natural Science Foundation of China Grants 21721003, Ministry of Science and Technology of China Grants 2016YFA0203200.  Hong Wang thanks for the help from professor Erkang Wang.

\appendix

\section{Derivations for $\sigma_{I}^{\pm}$ and $\langle\mathcal{H}\rangle$}
\label{sec:A}

The relationship between $\sigma_{I}^{\pm}$ and $\sigma^{\pm}$ is
\begin{equation}
\label{eq:A1}
\sigma_{I}^{\pm}=exp\big\{i\hat{H}_{grav}t\big\}\sigma^{\pm}exp\big\{-i\hat{H}_{grav}t\big\}.
\end{equation}
The form of $\sigma_{I}^{\pm}$ in \eqref{eq:A1} is complicated. And we need to simplify it. Bring \eqref{eq:4.63} into \eqref{eq:A1}, using \eqref{eq:4.59} and \eqref{eq:4.60}, then \eqref{eq:A1} becomes
\begin{eqnarray}\begin{split}
\label{eq:A2}
\sigma_{I}^{\pm}&=\sigma^{\pm}exp\Big\{\pm2i\alpha \Big((\sigma^{+})^{2}+(\sigma^{-})^{2}-2\Big)t\Big\}\\
                &=\sigma^{\pm}exp\big\{\pm2i \mathcal{\hat{H}}t\big\},
\end{split}
\end{eqnarray}
where
\begin{equation}
\label{eq:A3}
\alpha\equiv\frac{3}{16\gamma\sqrt{\triangle}},
\end{equation}
\begin{equation}
\label{eq:A4}
\mathcal{\hat{H}}\equiv \alpha\big((\sigma^{+})^{2}+(\sigma^{-})^{2}-2\big).
\end{equation}
From the definition \eqref{eq:A4}, we can show that the relationship between the operator $\mathcal{\hat{H}}$ and $\hat{H}_{grav}$ is given as
\begin{eqnarray}\begin{split}
\label{eq:A5}
\hat{H}_{grav}&=\frac{1}{2}(\mathcal{\hat{H}}|\mathcal{V}|+|\mathcal{V}|\mathcal{\hat{H}})\\&=\mathcal{\hat{H}}|\mathcal{V}|+o(\hbar).
\end{split}
\end{eqnarray}
In \eqref{eq:A5}, from the first step to the second step, we used a well-known conclusion which is the difference of different factor ordering is a higher order term of $\hbar$~\cite{EAJ}. Equation \eqref{eq:A5} shows that if we neglect the effect of $o(\hbar)$, then $\mathcal{\hat{H}}$ approximately represents the energy density operator and its eigenvalue represents the magnitude of the energy density.

\eqref{eq:A2} is still too complicated. To simplify the calculation, one can assume that the coupling constant $\xi$ is small. The interaction Hamiltonian is small compared with the Hamiltonians of the bath and the spacetime. In another word, the interaction Hamiltonian is micro while the Hamiltonian of the bath and the spacetime is macro. This is similar to the model which two macro systems coupled with a small interaction Hamiltonian. Thus in \eqref{eq:A2}, $\mathcal{\hat{H}}$ can be seen as an operator related to the macro system. Thus we introduce the semiclassical approximation: $\mathcal{\hat{H}}=\langle\mathcal{\hat{H}}\rangle+o(\hbar)$. Here, $\langle\mathcal{\hat{H}}\rangle$ represents the average value of the operator $\mathcal{\hat{H}}$. Therefore, neglecting $o(\hbar)$, equation \eqref{eq:A2} becomes
\begin{equation}
\label{eq:A6}
\sigma_{I}^{\pm}=\sigma^{\pm}exp\big\{\pm2i \langle\mathcal{\hat{H}}\rangle t\big\}.
\end{equation}
Combine equations \eqref{eq:A5}, \eqref{eq:4.6} and \eqref{eq:4.51}, we have
\begin{eqnarray}\begin{split}
\label{eq:A7}
\langle\mathcal{\hat{H}}\rangle&=2\pi\gamma\sqrt{\triangle}\cdot\langle\frac{\hat{H}_{grav}}{V}\rangle\\
                               &= -\frac{3}{4}\gamma \sqrt{\triangle}\cdot\frac{\dot{a}^{2}}{a^{2}}.
\end{split}
\end{eqnarray}

In \eqref{eq:A7}, from the first step to the second step, we approximately take the average value of the operator $\hat{H}_{grav}$ as the classical Hamiltonian  $H_{grav}$ in \eqref{eq:4.6}. Under this treatment, we have shown the gravitational subsystem can indeed reach the equilibrium state after the transient relaxation. This is a general feature that the system weekly coupled with a bath can reach the equilibrium steady state. It also indicates that the current treatment is reasonable.

As we limit the parameter $\xi$ to be a small constant, $H_{int}$ is small compared to $H_{grav}$ and $H_{\phi_{1}}$. This leads to  $H_{grav}+H_{\phi_{1}}= o(\xi)$. Taking the energy density of the scalar field as $\rho_{\phi_{1}}$, then we have
\begin{equation}
\label{eq:A8}
\frac{\dot{a}^{2}}{a^{2}}= \frac{8\pi}{3}\cdot \rho_{\phi_{1}}+o(\xi),
\end{equation}
this is the Friedmann equation. If the scalar field is in the thermal equilibrium state, it can be treated as a heat bath, the density matrix of the bath becomes~\cite{HC}
\begin{equation}
\label{eq:A9}
R_{0}=\prod_{\vec{k}}exp\Big(-\beta_{1}\omega_{k}A_{\vec{k}}^{\dag}A_{\vec{k}}\Big)\Big(1-exp(-\beta_{1}\omega_{k})\Big).
\end{equation}
Here, $\beta_{1}=1/T_{1}$ and $T_{1}$ is the temperature of the bath. Based on \eqref{eq:A9}, we can derive the mean particle number with the momentum $\vec{k}$ as~\cite{HC}
\begin{equation}
\label{eq:A10}
\bar{n}_{1}(\vec{k})=\mathrm{Tr}_{B}(R_{0}A_{\vec{k}}^{\dag}A_{\vec{k}})=\frac{e^{-\beta_{1}\omega_{k}}}{1-e^{-\beta_{1}\omega_{k}}}.
\end{equation}
The trace is taken over the bath. As the spin of the scalar particle is zero, it obeys the Bose-Einstein distribution. The total energy of the bath is (neglecting the vacuum energy)
\begin{equation}
\label{eq:A11}
E_{B}=\int d\vec{k}^{3}\,\omega_{k}\,\bar{n}_{1}(\vec{k}),
\end{equation}
The energy density $\rho_{\phi_{1}}$ of the scalar field is~\cite{SW}
\begin{eqnarray}\begin{split}
\label{eq:A12}
\rho_{\phi_{1}}&=\frac{E_{B}}{V}\\
    &=\frac{1}{V}\int d\vec{k}^{3}\,\omega_{k}\,\bar{n}_{1}(\vec{k})\\
    &\approx \frac{e^{-\beta_{1}m_{1}}}{\ell_{0}^{3}}\Big(64\pi m_{1}^{4}T_{1}^{3}+3\sqrt{2}\pi^{\frac{3}{2}}m_{1}^{\frac{3}{2}}T_{1}^{\frac{5}{2}}\Big).
\end{split}
\end{eqnarray}
In \eqref{eq:A12}, we approximately take $\bar{n}_{1}(\vec{k})= e^{-\beta_{1}\omega_{k}}$. This is reasonable when the bath is not under the low temperature. In addition, from the second step to the third step, we used the relation $\omega=\sqrt{\vec{k}^{2}/a^{2}+m_{1}^{2}}\approx m_{1}+\vec{k}^{2}/(2m_{1}a^{2})$ (non-relativistic approximation for the energy of the scalar particle) and $V=a^{3}\ell_{0}^{3}$.
Combining the equations \eqref{eq:A7}, \eqref{eq:A8} and \eqref{eq:A12}, we obtain
\begin{equation}
\label{eq:A13}
\langle\mathcal{H}\rangle= -\frac{\gamma \sqrt{\Delta}e^{-\beta_{1}m_{1}}}{\ell_{0}^{3}}\Big(128\pi^{2} m_{1}^{4}T_{1}^{3}+6\sqrt{2}\pi^{\frac{5}{2}}m_{1}^{\frac{3}{2}}T_{1}^{\frac{5}{2}}\Big).
\end{equation}

\end{document}